\documentclass[12pt,preprint]{aastex}

\usepackage{lscape}
\usepackage{textcomp}
\usepackage{graphicx}
\usepackage{epsfig}

\shorttitle{Chemical evolution of RMSs} \shortauthors{Naiping Yu
et al.}

\begin{document}


\title {Chemical evolution of Red MSX Sources in the southern sky}
\author
 {Naiping Yu\altaffilmark{1},Jinlong Xu\altaffilmark{1}}

\altaffiltext{1} {National Astronomical Observatories, Chinese
Academy of Sciences, Beijing 100012, China}


\label{firstpage}


\begin{abstract}

Red MSX  Sources(RMSs) are regarded as excellent candidates of
massive star-forming regions. In order to characterize the chemical
properties of massive star formation, we made a systematic study of
87 RMSs in the southern sky, using archival data taken from the
Atacama Pathfinder Experiment (APEX) Telescope Large Area Survey of
the Galaxy (ATLASGAL), the Australia Telescope Compact Array (ATCA),
and the Millimetre Astronomy Legacy Team Survey at 90 GHz (MALT90).
According to previous multi-wavelength observations, our sample
could be divided into two groups: massive young stellar objects
(MYSOs) and HII regions. Combined with the MALT90 data, we
calculated the column densities of N$_2$H$^+$, C$_2$H, HC$_3$N and
HNC, and found that they are not much different from previous
studies made in other massive star-forming regions. However, their
abundances are relatively low compared to infrared dark clouds
(IRDCs). The abundances of N$_2$H$^+$ and HNC in our sample are at
least one magnitude lower than those found IRDCs, indicating
chemical depletions in the relatively hot gas. Besides, the
fractional abundances of N$_2$H$^+$, C$_2$H and HC$_3$N seem to
decrease as a function of their Lyman continuum fluxes ($N_L$),
indicating these molecules could be destroyed by UV photons when HII
regions have formed inside. We also find the C$_2$H abundance
decreases faster than HC$_3$N with respect to $N_L$. The abundance
of HNC has a tight correlation with that of N$_2$H$^+$, indicating
it may be also preferentially formed in cold gas. We regard our RMSs
are in a relatively late evolutionary stage of massive star
formation.

\end{abstract}



\section{Introduction}
The study of massive stars are important to astrophysics. They
dominate the luminosity of galaxies. Their feedback to the
environment will have a profound effect on the physical and chemical
evolutions of the surrounding interstellar medium (ISM), and even
may trigger the next generation of star formation that will take
place therein (e.g. Yu et al. 2014). At the end of their short life,
they release heavy elements to the cosmic space in the form of
supernova explosions, which will have an immense impact to the
evolution of galaxies. In the last two decades, a lot of progress
has been made both in theories and observations of star formation.
However, massive star formation is still poorly understood compared
with their low-mass counterparts. One reason is that they are rare
and evolve quickly. The other reason is that they used to form in
dense clusters at far distances, which makes it hard to study them
individually. According to the review made by Zinnecker $\&$ Yorke
(2007), the processes of a massive star formation could roughly be
divided into four different observable stages. The first stage is
the prestellar phase that can be found in IRDCs (e.g. Perault et al.
1996; Egan et al. 1998; Peretto $\&$ Fuller 2009). The typical
temperature of this stage is about 10 K, as no stars have formed
inside. The next stage is the ``hot core'' phase, with various
complex organic molecular emissions inside and a typical temperature
of $\sim$ 100 K (e.g. Cesaroni et al. 1992; Cesaroni et al. 2010).
The third stage is hypercompact and ultracompact HII regions and the
final stage is compact and classical HII regions.

It is also important to understand chemical processes that may occur
during all stages of massive star formation. In the early stages of
massive star formation, the gas is cold and diffuse. Thus the
chemistry is dominated by low-temperature gas-phase reactions,
leading to the formation of small radicals and unsaturated molecules
(e.g. Herbst $\&$ Klemperer 1972). During the gas collapse phase,
the density of gas becomes so high and the temperature so low that
many atoms and molecules are absorbed onto the dust grains, forming
various kinds of icy mantles (e.g. Shematovich 2012). Many complex
reactions could take place between these mantles. These kind of
grain-surface chemical processes are great important to some organic
species. For example, both observations and chemical models indicate
methanol is mainly formed through grain-surface reactions, as the
production of methanol is ineffective by gas-phase reactions
(Wirstrom et al. 2011). The dust works as a catalyst in these
reactions. In the hot core stage, a massive young stellar object has
just formed inside. Its radiations quickly heat up the surrounding
gas and dust. Icy mantles on the dust evaporate back into the gas
phase. Warm gas-phase chemistry between these fresh material becomes
important both in the formation and destruction of many complex
organic molecules (e.g. Belloche et al. 2008; Qin et al. 2011). In
the HII regions, intense UV radiations are emitted from the central
O/B star(s), leading to the destructions and/or productions of many
species (e.g. Fuente et al. 1993; Yu $\&$ Wang 2015).

In order to study massive star formation, we first need a large
sample of massive star-forming candidates. By comparing the colors
of MSX and 2MASS point sources to those already known MYSOs, Lumsden
et al. (2002; 2013) selected about 2000 MYSO candidates. This is
known as the RMS survey. The RMS survey is still an on-going
project. Follow-up infrared and radio observations indicate they are
indeed excellent candidates of massive star-forming regions (e.g.
Urquhart et al. 2007; Cooper et al. 2013; Yu $\&$ Wang 2014).
However, chemical properties of RMSs are still less studied. What is
the chemistry in RMSs$?$ Are they similar to other massive
star-forming regions like IRDCs and/or extended green objects
(EGOs)$?$ With the aim to better understand the chemical evolution
of massive star formation, we have studied the chemical properties
of 87 RMSs and compared them with previous observations of other
massive star-forming regions in this paper.

\section{Data and Source Selections}
Our molecular line data comes from MALT90, which is an international
project aims to characterize the sites within our Galaxy where
massive star formation will take place (e.g. Foster et al. 2011;
Jackson et al. 2013). This project is performed with Mopra, a 22-m
single dish radio telescope located near Coonabarabran in New South
Wales, Australia. By splitting the Mopra Spectrometer (MOPS) into 16
zoom bands of 138 MHz each, Mopra could map 16 molecular lines
simultaneously with a velocity resolution of $\sim$ 0.11 km s$^{-1}$
at frequencies near 90 GHz. The critical densities of these
molecular lines range from 2 $\times$ 10$^5$ to 3 $\times$ 10$^6$
cm$^{-3}$ (see Table 1 in Rathborne et al. 2014). And these lines
will help us to probe the physical, chemical, and evolutionary
states of dense high-mass star-forming clumps. The angular
resolution of Mopra is 38$^{\prime\prime}$, with a beam efficiency
between 0.49 at 86 GHz and 0.42 at 115 GHz (Ladd et al. 2005). The
pointing accuracy of the main MALT90 maps is about
8$^{\prime\prime}$ and the absolute flux uncertainty is in the range
of 10 $\sim$ 17$\%$ depending on the transition in question (Foster
et al. 2013). The target clumps of this survey are selected from the
870 micron APEX Telescope Large Area Survey of the Galaxy (ATLASGAL)
(Schuller et al. 2009; Contreras et al. 2013), with Galactic
longitude range of $\sim$ -60$^{\circ}$ to $\sim$ 15$^{\circ}$ and
Galactic latitude range of -1$^{\circ}$ to +1 $^{\circ}$. The MALT90
data are originally stored in RPFITS format. Using software packages
of CLASS (Continuum and Line Analysis Single-Disk Software) and GREG
(Grenoble Graphic), we conducted the data. The data files are
publicly available and can be downloaded from the MALT90 Home
Page\footnote{ http://atoa.atnf.csiro.au/MALT90}.

Using ATCA, Urquhart et al. (2007) observed radio emissions of 826
RMSs in the southern sky. This programme divided their sources into
three groups: MYSO candidates, HII regions and others such as
evolved stars and planetary nebulae (PNe). Our sources are selected
from this sample. Considering radio interferometers may have
difficulty imaging extended emissions in the Galactic plane, we also
have checked our sources with the data taken from the Sydney
University Molonglo Sky Survey (SUMSS) carried out at 843 MHz with
the Molonglo Observatory Synthesis Telescope (MOST) (Mauch et al.
2003). In our sample, sources are regarded as MYSO candidates if no
radio emissions have been detected by ATCA and MOST, otherwise they
are regarded as HII regions. We should mention here that even though
many RMSs are radio-quiet in the observations of ATCA and MOST, they
are still regarded as HII region candidates in the Red MSX Source
survey database\footnote{
http://rms.leeds.ac.uk/cgi-bin/public/RMS$_{-}$DATABASE.cgi},
because their positions in the color-color diagrams (CCD) are
consistent with already known HII regions. These sources may be real
MYSOs or HII regions that their radio emissions are too weak to be
detected. Deeper and higher spacial resolution observations should
be carried out in the future to found out whether HII regions have
formed inside these sources. In order to make our sample
unambiguous, these sources are not involved in our study. Besides,
given the large beam size of Mopra, the effective diameter of our
RMS clumps given by Contreras et al. (2013) should be larger than
38$^{\prime\prime}$ and far from known bubbles or HII regions.
Finally, we found 87 RMSs (28 MYSOs and 59 HII regions) having
molecular data in MALT90. Our source selection criteria could be
summarized in the followings: (i) Sources should be relatively
isolated and be detected by MALT90; (ii) The effective diameter of a
RMS clump should be larger than 38$^{\prime\prime}$; (iii) For
MYSOs, they should be radio-quiet and have similar near and mid-IR
colors to already known MYSOs. (iv)For HII regions, radio emissions
should have been detected by ATCA and/or MOST, and have similar near
and mid-IR colors to already known HII regions. The basic
information of our sources is listed in Table 1.

\section{Results}
\subsection{Dust Temperature and $H_2$ Column Densities}
Dust temperature ($T_d$) is essential in the study of chemical
evolution in star formation regions. When $T_d$ is below $\sim$ 20
K, carbon species like CO and CS can be depleted in the cold gas.
Derived from adjusting single-temperature dust emission models to
the far-infrared intensity maps measured between 160 and 870 $\mu$m
from the Herschel and APEX sky surveys, Guzm\'{a}n et al. (2015)
recently have calculated the dust temperatures and $H_2$ column
densities for $\sim$ 3000 MALT90 clumps. The FWHM of these data
differs from 12$^{\prime\prime}$ to 35$^{\prime\prime}$ (the data
from the PACS 70 $\mu$m band is excluded). To make an adequate
comparison, they convolved all images to a spacial resolution of
35$^{\prime\prime}$, which is the lowest resolution given by the 500
$\mu$m SPIRE instrument. We used these dust temperatures and $H_2$
column densities derived by Guzm\'{a}n et al. (2015) in our study
due to the similar beam resolution of Mopra (35$^{\prime\prime}$ vs
38$^{\prime\prime}$). These values are also listed in Table 1
(column 3 and column 5) and the dust temperature distributions are
shown in Fig. 1. The mean dust temperature of our RMSs is 26.4 K.
This is consistent with many other observations of MYSOs and HII
regions (e.g. Hennemann et al. 2009; Sreenilayam $\&$ Fich 2011).
However, our values are higher than those found in IRDCs (e.g.
Vasyunina et al. 2011; Hoq et al. 2013), indicating our sources are
relatively more evolved. From Fig. 1, it can also be noted that in
our two subgroups, HII regions are warmer than MYSOs on average.
Kolmogorov-Smirnov (K-S) test gives a probability of less than 0.01
$\%$ that these two groups originate from the same parent
population. Typically, the difference will be regarded as
significant when the K-S test gives a percentage lower than 5$\%$.
It is reasonable that gas and dust will get warmer as the central
star(s) evolves.

Urquhart et al. (2007) have completed the 4.8 and 8.6 GHz
observations of 826 RMS sources in the southern sky using
observations of ATCA. We use these data to estimate the Lyman
continuum fluxes of our HII regions. Assuming an electron
temperature of $T_e$ = 10$^4$ K, the number of UV ionizing photons
needed to keep an HII region ionized could be given as (Chaisson
1976; Guzm\'{a}n et al. 2012)
\begin{equation}
N_L = 7.6 \times 10^{46} (\frac{S_\nu}{Jy}) (\frac{D}{kpc})^2
(\frac{\nu}{GHz})^{0.1} (\frac{T_e}{10^4 K})^{-0.45} s^{-1}
\end{equation}
where $\nu$ is the frequency and $S_\nu$ is the integrated flux
density given by Urquhart et al. (2007), $D$ is the kinematic
distance derived by Urquhart et al. (2008). The large number of
Lyman continuum fluxes (listed in Table 1, column 4) indicate that
they are probably massive O or early B-type star formation regions.

\subsection{Column Densities and Abundances}
Assuming local thermodynamic equilibrium (LTE) conditions and a beam
filling factor of 1, the column densities of N$_2$H$^+$, C$_2$H,
HC$_3$N and HNC could be derived from (e.g. Garden et al. 1991;
Sanhueza et al. 2012):
\begin{equation}
N = \frac{8 \pi \nu^3}{c^3 R} \frac{Q_{rot}}{g_u A_{ul}}
\frac{exp(E_l/k T_{ex})}{1 - exp (-h \nu /k T_{ex})} \frac{\tau}{1
- exp(-\tau)} \frac{\int T_{mb} dv}{J(T_{ex}) - J(T_{bg})}
\end{equation}
where $c$ is the velocity of light in the vacuum, $g_u$ is the
statistical weight of the upper level, $A_{ul}$ is the Einstein
coefficient for spontaneous transition, $E_l$ is the energy of the
lower level, $Q_{rot}$ is the partition function, $T_{bg}$ is the
background temperature, $\tau$ is the optical depth, $T_{ex}$ is the
excitation temperature in all cases. We assume that $T_{ex}$ is
equal to the dust temperature ($T_d$) derived by Guzm\'{a}n et al.
(2015). $R$ is only relevant for hyperfine transitions because it
takes into account the satellite lines correcting by their relative
opacities. The value of $R$ is 5/9 for N$_2$H$^+$, 5/12 for C$_2$H
and 1.0 for transitions without hyperfine structure. The values of
$g_u$, $A_{ul}$ and $E_l$ could be found in the Cologne Database for
Molecular Spectroscopy (CDMS) (M\"{u}ller et al. 2001, 2005 ).
$J(T)$ is defined by
\begin{equation}
J(T) = \frac{h \nu}{k} \frac{1}{e^{h \nu/k T} - 1}
\end{equation}

Near 90 GHz, the N$_2$H$^+$ (1-0) line has 15 hyperfine structures
(HFS) out of which seven have a different frequency (e.g. Pagani et
al. 2009; Keto $\&$ Rybicki 2010). In massive star-forming regions,
these velocity components tend to blend into three groups because of
turbulence (see Fig. 2 of Purcell et al. 2009). We follow the
procedure outlined by Purcell et al. (2009) to estimate the optical
depth of N$_2$H$^+$. Assuming the line widths of the individual
hyperfine components are all equal, the integrated intensities of
the three blended groups should be in the ratio of 1:5:3 under
optically thin conditions. The optical depth can then be derived
from the ratio of the integrated intensities of any two groups using
the following equation:
\begin{equation}
\frac{\int T_{MB,1} dv}{\int T_{MB,2} dv} = \frac{1 -
exp(-\tau_1)}{1 - exp(-a\tau_1)}
\end{equation}
where $a$ is the expected ratio of $\tau_2/\tau_1$ under optically
thin conditions. We determined the optical depth only from the
intensity ratio of group 1/group 2 (defined by Purcell et al. 2009),
as anomalous excitation of the $F_1F$ = 10-11 and 12-12 components
(in our group 3) has been reported by Caselli et al. (1995). Here
the value of $a$ is 5. To derive the line intensities and peak
emissions, we fitted the three groups with three Gaussian profiles
from the averaged pixels inside 38$^{\prime\prime}$. The integrated
intensities of group 1 and group 2 are listed in table 2 and the
derived column densities of N$_2$H$^+$ are listed in table 4.
Finally, the fractional abundance of N$_2$H$^+$ with respect to
H$_2$ could be estimated by $\chi$ (N$_2$H$^+$) =
N(N$_2$H$^+$)/N(H$_2$). The derived values are listed in Table 5.

C$_2$H ($N = 1 - 0$) has six hyperfine components out of which two
($N = 1 - 0, J = 3/2 - 1/2, F = 2 - 1$ and $N = 1 - 0, J = 3/2 -
1/2, F = 1 - 0$) could easily be detected in MALT90. The optical
depth of C$_2$H ($F=2-1$) can be obtained by comparing its hyperfine
components. Under the assumption of optically thin, the intensity
ratio of C$_2$H ($F=2-1$) and C$_2$H ($F=1-0$) should be 2.0 (Tucker
et al. 1974). Thus, the opacity of the brightest component of C$_2$H
could be given by
\begin{equation}
\frac{1 - e^{-0.5\tau}}{1 - e^{-\tau}} =
\frac{T_{mb}(F=1-0)}{T_{mb}(F=2-1)}
\end{equation}
The peak intensities of these two components of C$_2$H are listed in
table 3. The column densities of C$_2$H can be derived through
equation 2. The calculated parameters are also listed in table 4.

Many studies indicate HC$_3$N ($J=10-9$) used to be optically thin
(e.g. Chen et al. 2013). In order to compare our study with that of
Vasyunina et al. (2011) and Sanhueza et al. (2012), we also assume
that this transition is optically thin here. The optical thickness
of HNC could be estimated by comparing the intensity of its
isotopologue HN$^{13}$C:
\begin{equation}
\frac{1 - e^{-\tau_{12}}}{1 - e^{-\tau_{12}/X}} =
\frac{^{12}T_{mb}}{^{13}T_{mb}}
\end{equation}
where $X$ $\sim$ $[^{12}C]/[^{13}C]$ is the isotope abundance ratio.
Here we use a constant $X$ = 50 in our calculations (Purcell et al.
2006).

\section{Analysis and Discussions}
\subsection{N$_2$H$^+$ ($Diazenylium$)}
N$_2$H$^+$ is one of the most detected molecules in our study.
According to previous observations made in low-mass star-forming
regions and chemical evolution models, N$_2$H$^+$ is mainly
destroyed via N$_2$H$^+$ + CO $\rightarrow$ HCO$^+$ + N$_2$ when CO
evaporates into the gas phase (e.g. Bergin $\&$ Langer 1997; Lee et
al. 2004). Thus we would expect to find that the N$_2$H$^+$
abundance decreases as a function of evolutionary stage. However,
this trend was not found in the observations of Sanhueza et al.
(2012) and Hoq et al. (2013). They regard this may be caused by
their beam dilution, and/or chemical processes of N$_2$H$^+$ in
massive star-forming regions may be different from those in low-mass
star-forming regions. In our study, we find the column densities of
N$_2$H$^+$ range from 0.31 $\times$ 10$^{13}$ cm$^{-2}$ to 4.24
$\times$ 10$^{13}$ cm$^{-2}$, with a mean value of 1.51 $\times$
10$^{13}$ cm$^{-2}$. These values are not so much different from
those found in other massive star-forming regions like IRDCs and
EGOs (see table 6). However, we find our N$_2$H$^+$ abundances range
from 0.4 $\times$ 10$^{-10}$ to 0.8 $\times$ 10$^{-9}$, with a mean
value of only 0.2 $\times$ 10$^{-9}$. These values are at least one
magnitude lower than those found in IRDCs. This may be because our
RMSs are relatively more evolved than IRDCs. Fig.2 shows the
histograms of the number distributions of N$_2$H$^+$ abundances for
our total RMSs, MYSOs and HII regions. K-S test gives a possibility
of 49$\%$ that the two distributions originate from the same parent
population, indicating that they are undistinguishable. We found the
fractional abundance of N$_2$H$^+$ ($\chi$(N$_2$H$^+$)) seems to
decrease as a function of Lyman continuum fluxes $N_L$ as shown in
Fig. 3. This may be caused by that in HII regions, N$_2$H$^+$ could
be destroyed by recombination with electron : N$_2$H$^+$ + e$^-$
$\rightarrow$ N$_2$ + H or NH + N (e.g. Dislaire et al. 2012; Vigren
et al. 2012; Yu et al. 2015).

\subsection{C$_2$H ($Ethynyl$)}
C$_2$H was first detected by Tucker et al. (1974) in interstellar
clouds. However it has not been systematically studied and its
evolution routes are still not clear in massive star-forming
regions. Through observations of NGC7023, Fuente et al. (1993)
suggest C$_2$H is a good photodissociation region (PDR) tracer. At
the boundary layers between ionized and molecular gas, UV photons
can produce C$_2$H by reactions with acetylene (C$_2$H$_2$):
C$_2$H$_2$ + $h\nu$ $\rightarrow$ C$_2$H + H. However, based on the
high spacial resolution observations of the Submillimeter Array
(SMA) and chemical models, Beuther et al. (2008) suggest C$_2$H may
preferentially exist in the outer part of dense gas clumps, as they
found that the distribution of C$_2$H shows a hole in the hot core.
In the cold low-mass star-forming cores of Taurus, Pratap et al.
(1997) also found strong C$_2$H emissions. In the high-mass regime
of the prestellar IRDC G028.23-00.19, Sanhueza et al. (2013)
interestingly found that C$_2$H have similar spatial distribution
and line widths than NH$_2$D, suggesting that C$_2$H can also trace
cold gas in IRDCs. Thus, C$_2$H may trace both early and late stages
of star formation. Recently, Li et al. (2012) observed 27 massive
star-forming regions with water masers. They found their C$_2$H
optical depth declines when molecular clouds evolve from
hypercompact HII regions to classical HII regions. On the other
hand, Yu $\&$ Wang (2015) found that the [C$_2$H]/[H$^{13}$CO$^+$]
relative abundance ratio decreases from MYSOs to HII regions. These
studies suggest that C$_2$H might be used as a chemical clock for
molecular clouds. In this study, the abundance of C$_2$H does not
show significant difference in our subgroups of MYSOs and HII
regions (see Fig. 4). The K-S test gives a probability of 61$\%$
that they may originate from the same parent population. The
abundances of C$_2$H range from 1.4 $\times$ 10$^{-9}$ to 1.96
$\times$ 10$^{-8}$, with an average value of 0.50 $\times$
10$^{-8}$. Our mean value is about 3 to 7 times less than those
found by Sanhueza et al. (2012) and Vasyunina et al. (2011),
indicating slight depletion of C$_2$H. Fig. 3 (the middle panel)
shows the C$_2$H fractional abundance plotted as a function of $N_L$
in logarithmic scales. It seems the fractional abundance of C$_2$H
decreases as the number of UV photons increases. One reason may be
that in dense molecular gas, nearly all carbon is in the form of CO
because CO is chemically stable. However, in HII regions UV photons
can photodissociate CO into C and O (CO + $h\nu$ $\rightarrow$ C +
O). Atom C can further be photoionized into C$^+$ (C + $h\nu$
$\rightarrow$ C$^+$ + e$^-$). The enhanced abundances of O and C$^+$
can destroy C$_2$H through reactions C$_2$H + O $\rightarrow$ CO +
CH and C$_2$H + C$^+$ $\rightarrow$ C$_3$$^+$ + H (Watt et al.
1988). Deep observations and chemical models should be carried out
to verify our speculations.

\subsection{HC$_3$N ($Cyanoacetylene$)}
HC$_3$N traces both cold molecular clouds and hot cores. Previous
studies indicate HC$_3$N is mainly produced through C$_2$H$_2$ + CN
$\rightarrow$ HC$_3$N + H (Chapman et al. 2009). It is the most
simple example of the cyanopolyynes, HC$_{2n+1}$N. The HC$_3$N
abundances we derived range from 0.4 $\times$ 10$^{-10}$ to 1.6
$\times$ 10$^{-10}$, with a mean value of 0.9 $\times$ 10$^{-10}$
(see table 6). These values are about 5 times less than previous
studies made in IRDCs by Sanhueza et al. (2012) and Vasyunina et al.
(2011). Fig.5 also shows the number distributions of the HC$_3$N
abundances for MYSOs and HII regions. No real change is clear. The
K-S test gives a probability of 59$\%$ that they may originate from
the same parent population. The bottom panel of Fig. 3 shows the
relationship between $\chi$(HC$_3$N) and $N_L$. It also indicates
that the abundance of HC$_3$N decreases when increasing UV field.
One reason maybe that in the strong UV field of HII regions, the
reaction C$_2$H$_2$ + $h\nu$ $\rightarrow$ C$_2$H + H dominates
instead, leading to the production of HC$_3$N ineffective. However,
if this scenario is true, it means that C$_2$H$_2$ tends to form
more C$_2$H over HC$_3$N. Fig. 6 shows the [C$_2$H]/[HC$_3$N]
relative abundance rations plotted as a function of $N_L$ in
logarithmic scales. It can be noted that the HC$_3$N abundance does
not decrease faster than C$_2$H with respect to $N_L$. There must be
another scenario for the destruction of HC$_3$N in HII regions. More
deep observations and chemical models should be carried out to find
this scenario.

\subsection{HNC and HN$^{13}$C ($Hydrogen$ $Isocyanide$)}
HNC and its isotopologue HN$^{13}$C are prevalent in cold gas (e.g.
Bergin $\&$ Tafalla 2007). Hirota et al. (1998) found abundances of
HNC in OMC-1 are 1 $\sim$ 2 orders of magnitude less than those in
dark cloud cores, indicating destruction of HNC in hot regions.
Sanhueza et al (2012) also found the N$_2$H$^+$/HNC abundance ratio
slightly decrease as IRDCs evolve (see their Fig. 17), suggesting
HNC may be preferentially formed in cold gas. In our study, we found
the HN$^{13}$C detection rate in MYSOs is about 2 times higher than
in HII regions (39$\%$ vs 20$\%$). The HNC abundances we derived in
our all RMSs range from 0.5 $\times$ 10$^{-9}$ to 5.6 $\times$
10$^{-9}$, with a mean value of 0.2 $\times$ 10$^{-8}$. This mean
value is about 20 times lower than those found in IRDCs by Sanhueza
et al. (2012), which would be consistent with HNC being
preferentially formed in the cold gas. Although our mean value is
similar to that of Vasyunina et al. (2011), we should mention that
in their study, Vasyunina et al. (2011) made an assumption that the
emission of HNC is optically thin. This assumption will make column
densities at least one order of magnitude lower. We also find a
tight correlation between $\chi$(N$_2$H$^+$) and $\chi$(HNC) (Fig.
7). The functional form of the linear fit is:
\begin{equation}
\chi(HNC) =1.44 \times \chi(N_2H^+)
\end{equation}
with a correlation coefficient $r$ = 0.74. Our study also support
HNC and HN$^{13}$C are dense cold tracers.

\subsection{Why undistinguishable ?}
RMSs are regarded as excellent candidates of massive star-forming
regions. Multiwavelength observations have been carried out to study
their physical parameters. However their chemical properties are
less studied. In this paper, we selected 87 RMSs to study their
chemical properties. We found the column densities of N$_2$H$^+$,
C$_2$H, HC$_3$N and HNC are not much different from previous
observations made in IRDCs and EGOs. However, their abundances are
relatively low, indicating depletions of them in the late stages of
massive star formation, especially those of N$_2$H$^+$ and HNC. The
value of $\chi$ (N$_2$H$^+$) and $\chi$ (HNC) in our sample are at
least one magnitude lower than those found in IRDCs. This may be
because that our RMSs are relatively more evolved. In our two
subgroups of MYSOs and HII regions, K-S tests suggest that the
differences of their chemical properties are not significant. Two
reasons may be responsible for this. Firstly, the angular resolution
of our MALT90 data is not high enough. At 2.9 kpc, the
38$^{\prime\prime}$ beam dilution corresponds to over half a parsec.
Therefore, MALT90 probes not only star-forming cores, but also
diffuse envelopes around. Secondly, our source selections are mainly
based on the work of Lumsden et al. (2002; 2013) and Urquhart et al.
(2007). On the other hand, our sources are also involved in the work
of Guzm\'{a}n et al. (2015). By combining other sky survey data
taken with the $Herschel$ and $Spitzer$ telescope, Guzm\'{a}n et al.
(2015) divided their sample into ``Quiescent'', ``Protostellar'',
``HII region'' and ``Photodissociation region (PDR)'' clumps. 83$\%$
(49 in 59) of our HII regions are consistent with their
classifications. This means that our source selection criteria for
HII regions are reliable. However, according to their
classification, our 28 MYSOs involves 19 protostellar and 9 HII
regions. Even though all of our MYSO candidates are radio-quiet and
have similar infrared colors to already known MYSOs, most of them
also show bright emissions at 8 $\mu$m. This suggests that our MYSO
candidates are probably in the late stages of protostellar and HII
regions may form inside in the near future.

\section{Conclusions}
We have studied chemical properties of 87 RMSs in the southern sky,
using archival data taken from APEX, ATCA, and Mopra. According to
previous multiwavelength observations, we divide our sample into two
groups: MYSOs and HII regions. We find the column densities of
N$_2$H$^+$, C$_2$H, HC$_3$N and HNC are not much different from
previous studies made in other massive star-forming regions.
However, their abundances are relatively low compared to IRDCs. The
abundances of N$_2$H$^+$ and HNC in our sample are at least one
magnitude lower than those found IRDCs, indicating chemical
depletions in the relatively hot gas. The fractional abundances of
N$_2$H$^+$, C$_2$H and HC$_3$N seem to decrease as a function of
their Lyman continuum fluxes, indicating these molecules could be
destroyed when HII regions have formed. Besides, the C$_2$H
abundance seems to decrease faster than HC$_3$N in HII regions. The
abundance of HNC has a tight correlation with that of N$_2$H$^+$,
indicating it may be also preferentially formed in cold gas. We
regard our RMSs are in a relatively late evolutionary stage of
massive star formation.

\section{ACKNOWLEDGEMENTS}
We thank the referee for detailed comments which have considerably
improved this paper. This paper has made use of information from the
Red MSX Source survey database
http://rms.leeds.ac.uk/cgi-bin/public/RMS$_{-}$DATABASE.cgi and the
ATLASGAL Database Server
http://atlasgal.mpifr-bonn.mpg.de/cgi-bin/ATLASGAL$_-$DATABASE.cgi.
The Red MSX Source survey was constructed with support from the
Science and Technology Facilities Council of the UK. The ATLASGAL
project is a collaboration between the Max-Planck-Gesellschaft, the
European Southern Observatory (ESO) and the Universidad de Chile.
This research made use of data products from the Millimetre
Astronomy Legacy Team 90 GHz (MALT90) survey. The Mopra telescope is
part of the Australia Telescope and is funded by the Commonwealth of
Australia for operation as National Facility managed by CSIRO. This
paper is supported by National Natural Science Foundation of China
under grants of 11503037.

\appendix

\tabletypesize{\scriptsize}

\setlength{\tabcolsep}{0.1in}

\begin{deluxetable}{lcccr}
\tablewidth{0pt} \tablecaption{List of our sources.}
\renewcommand{\arraystretch}{1.2}
\tablehead{
 RMS   & $D$$^a$    & $T_d$$^b$  & log($N_L$) &  $N(H_2)$$^b$\\
 name    &(kpc) & (K)    & (S$^{-1}$)   &($\times$ 10$^{22}$ cm$^{-2}$)
}\startdata

    &  & HII regions\\
    \hline\\
G300.5047-00.1745  & 8.9  &27.2 (0.7  )     & ...         & 7.85  (0.36  ) \\

G301.1364-00.2249  & 4.3  &29.0 (1.0  )     & 47.31       & 44.12 (3.09  ) \\

G301.8147+00.7808  & 4.4  &29.0 (3.0  )     & 46.30       & 3.12  (0.37  ) \\
G302.0319-00.0613  & 4.5  &34.0 (1.0  )     & ...         & 4.51  (0.32  ) \\
G302.4867-00.0308  & 4.5  &27.0 (4.0  )     & 46.12       & 4.41  (0.61  ) \\

G307.5606-00.5871  & 7.4  &27.1 (0.7  )     & ...         & 13.32 (0.62  )  \\
G307.6138-00.2559  & 7.0  &29.0 (1.0  )     & ...         & 5.68  (0.40  )  \\

G309.9206+00.4790  & 5.4  &33.0 (2.0  )     & 47.40       & 12.43 (1.11  ) \\
G310.1420+00.7583  & 5.4  &29.0 (1.0  )     & ...         & 10.34 (0.72  ) \\
G311.6264+00.2897  & 7.3  &31.0 (1.0  )     & 47.40       & 13.63 (0.95  ) \\
G316.1386-00.5009  & 7.7  &27.0 (1.0  )     & ...         & 7.32  (0.51  )  \\
G317.4298-00.5612  &14.2  &34.0 (2.0  )     & 46.65       & 3.93  (0.35  )  \\
G318.9148-00.1647  &11.0  &34.0 (2.0  )     & 48.73       & 6.68  (0.80  ) \\

G320.7779+00.2412  &12.3  &25.0 (2.0  )     & 46.72       & 3.50  (0.42  ) \\
G321.7209+01.1711  & 2.8  &27.0 (3.0  )     & ...         & 11.08 (1.32  ) \\
G323.4468+00.0968  & 4.1  &24.0 (0.6  )     & ...         & 4.95  (0.35  ) \\

G324.1997+00.1192  & 6.8  &36.0 (2.0  )     & 48.32       & 15.65 (1.40  ) \\
G326.4477-00.7485  & 4.0  &24.0 (1.0  )     & 45.41       & 7.67  (0.54  ) \\
G326.4719-00.3777  & 3.4  &28.0 (1.0  )     & 47.02       & 10.83 (0.76  ) \\
G326.7249+00.6159  & 1.8  &31.0 (1.0  )     & 47.09       & 15.65 (1.10  ) \\
G328.3067+00.4308  & 5.1  &40.0 (1.0  )     & ...         & 13.32 (0.93  ) \\
G328.8074+00.6324  & 2.8  &25.0 (2.0  )     & ...         & 58.16 (5.23  )\\
G328.9580+00.5671  & 7.2  &26.0 (1.0  )     & 45.51       & 7.15  (0.50  ) \\
G329.3371+00.1469  & 7.2  &38.0 (1.0  )     & ...         & 12.43 (0.87  ) \\
G329.4055-00.4574  & 4.3  &27.0 (2.0  )     & ...         & 14.95 (1.79  ) \\
G329.4211-00.1631  & 4.5  &25.0 (2.0  )     & ...         & 6.68  (0.60  ) \\
G329.4761+00.8414  & 4.8  &30.0 (0.8  )     & 46.77       & 3.05  (0.14  ) \\
G329.8145+00.1411  & 4.9  &23.0 (0.6  )     & 45.27       & 6.83  (0.48  ) \\
G330.2845+00.4933  & 5.3  &25.0 (1.0  )     & 46.40       & 4.73  (0.33  ) \\
G330.2935-00.3946  &10.0  &32.0 (1.0  )     & 48.18       & 10.83 (0.76  ) \\
G330.8845-00.3721  & 3.8  &28.0 (1.0  )     & 47.22       & 39.32 (2.75  )\\
G330.9544-00.1817  & 9.6  &37.0 (1.0  )     & 48.63       & 66.77 (4.67  ) \\
G331.1282-00.2436  & 4.9  &24.4 (0.6  )     & ...         & 31.96 (1.50  ) \\
G331.4181-00.3546  & 3.9  &23.0 (5.0  )     & 46.54       & 9.22  (0.81  ) \\
G331.5582-00.1206  & 5.5  &26.1 (0.8  )     & ...         & 18.82 (1.32  ) \\
G332.1544-00.4487  & 3.6  &37.2 (0.8  )     & ...         & 7.49  (0.35  ) \\
G332.2944-00.0962  & 3.6  &24.0 (2.0  )     & 46.56       & 17.56 (1.58  ) \\
G332.8256-00.5498  & 3.6  &30.0 (3.0  )     & 47.84       & 60.90 (7.30  )\\
G333.0299-00.0645  & 3.6  &29.0 (3.0  )     & ...         & 4.62  (0.64  ) \\
G333.0494+00.0324  & 3.6  &27.0 (2.0  )     & ...         & 3.84  (0.34  ) \\
G333.1642-00.0994  & 5.1  &27.0 (2.0  )     & ...         & 4.95  (0.59  ) \\
G336.3684-00.0033  & 7.7  &23.0 (1.0  )     & 47.96       & 15.65 (1.10  )  \\
G337.0047+00.3226  &11.4  &26.0 (3.0  )     & 47.93       & 5.18  (0.62  ) \\
G337.4050-00.4071  & 3.1  &27.5 (0.7  )     & 46.53       & 38.42 (1.80  )\\
G337.8442-00.3748  & 3.0  &29.9 (0.9  )     & 45.77       & 6.23  (0.44  ) \\
G337.9947+00.0770  & 7.8  &25.7 (0.7  )     & ...         & 5.82  (0.41  ) \\
G338.0715+00.0126  & 3.0  &25.8 (0.9  )     & ...         & 16.77 (1.17  ) \\

G339.1052+00.1490  & 4.7  &25.0 (1.0  )     & 46.38       & 4.73  (0.33  ) \\
G340.0708+00.9267  & 4.6  &26.0 (2.0  )     & 46.73       & 5.95  (0.53  ) \\
G340.2480-00.3725  & 3.7  &22.0 (1.0  )     & 46.75       & 14.28 (1.00  ) \\
G342.7057+00.1260  & 3.4  &25.2 (0.7  )     & 45.18       & 15.30 (1.07  ) \\
G344.4257+00.0451  & 4.7  &30.9 (0.8  )     & 47.20       & 8.60  (0.60  ) \\
G345.0034-00.2240  & 2.8  &25.8 (11.0 )     & ...         & 41.17 (10.29 )\\
G345.4881+00.3148  & 2.1  &28.0 (1.0  )     & 47.41       & 35.86 (2.51  )\\
G345.6495+00.0084  &14.8  &35.0 (0.6  )     & 49.09       & 11.34 (0.79  )  \\
G346.0774-00.0562  &10.9  &24.0 (4.0  )     & 46.70       & 6.09  (0.85  ) \\
G346.5235+00.0839  &11.2  &29.0 (2.0  )     & 47.73       & 4.41  (0.52  ) \\
G347.8707+00.0146  & 3.4  &26.0 (1.0  )     & 46.84       & 7.32  (0.65  )  \\
G348.8922-00.1787  &11.2  &30.0 (2.0  )     & ...         & 5.55  (0.49  ) \\
\hline\\
& &  MYSOs\\
\hline\\
G305.2017+00.2072  &4.0   & 26.8 (0.8 )    & ...          &31.96(2.24 )  \\
G310.0135+00.3892  &3.2   & 28.0 (1.0 )    & ...          &8.21 (0.58 )  \\
G313.7654-00.8620  &7.8   & 23.0 (2.0 )    & ...          &11.87(1.06 )  \\
G314.3197+00.1125  &3.6   & 22.0 (1.0 )    & ...          &7.15 (0.50 )  \\
G318.9480-00.1969  &2.4   & 25.6 (0.4 )    & ...          &17.16(1.20 )  \\
G320.2437-00.5619  &9.5   & 18.0 (2.0 )    & ...          &5.95 (0.53 )   \\
G326.4755+00.6947  &1.8   & 22.3 (0.6 )    & ...          &38.42(2.69 )  \\
G326.5437+00.1684  &4.4   & 19.0 (4.0 )    & ...          &10.11(1.70 )  \\
G327.1192+00.5103  &4.9   & 27.0 (1.0 )    & ...          &7.85 (0.55 )  \\
G329.0663-00.3081  &11.6  & 18.5 (0.6 )    & ...          &13.02(0.91 )    \\
G330.9288-00.4070  &11.9  & 20.5 (0.5 )    & ...          &9.22 (0.65 )  \\
G332.4683-00.5228  &3.6   & 21.0 (1.0 )    & ...          &9.22 (0.65 )  \\
G332.9636-00.6800  &3.2   & 22.0 (1.0 )    & ...          &21.61(1.51 )  \\
G333.3151+00.1053  &3.6   & 23.0 (1.0 )    & ...          &9.43 (0.84 )  \\
G333.3752-00.2015  &3.6   & 23.0 (2.0 )    & ...          &4.73 (0.42 )  \\
G335.0611-00.4261  &2.8   & 22.5 (1.0 )    & ...          &14.28(1.00 )  \\
G337.1555-00.3951  &3.1   & 17.6 (0.5 )    & ...          &6.68 (0.47 )  \\
G338.2801+00.5419  &4.1   & 20.8 (0.8 )    & ...          &7.15 (0.50 )  \\
G338.9196+00.5495  &4.2   & 20.0 (1.0 )    & ...          &58.16(4.07 ) \\
G339.6221-00.1209  &2.8   & 25.6 (1.0 )    & ...          &7.67 (0.54 )  \\
G341.1281-00.3466  &3.3   & 23.3 (0.4 )    & ...          &6.83 (0.48 )  \\
G341.2182-00.2136  &3.4   & 23.7 (0.6 )    & ...          &12.72(0.59 )  \\
G342.9583-00.3180  &12.7  & 21.2 (0.5 )    & ...          &3.42 (0.16 )    \\
G343.5213-00.5171  &3.2   & 20.0 (2.0 )    & ...          &6.53 (0.78 )  \\
G345.2619-00.4188  &2.7   & 19.9 (0.5 )    & ...          &4.62 (0.21 )  \\
G345.7172+00.8166  &1.6   & 20.0 (1.0 )    & ...          &11.60(0.81 )   \\
G346.4809+00.1320  &15.0  & 23.0 (2.0 )    & ...          &2.91 (0.34 )  \\
G348.6491+00.0225  &11.1  & 21.0 (1.0 )    & ...          &4.95 (0.44 )   \\
\enddata

\tablenotetext{a} {Urquhart et al. (2008).}

\tablenotetext{b}{These values are from Guzm\'{a}n et al. (2015).}

\end{deluxetable}

\clearpage


\tabletypesize{\scriptsize}

\setlength{\tabcolsep}{0.1in}

\begin{deluxetable}{lrrrrrr}
\tablewidth{0pt} \tablecaption{ Integrated intensities of
N$_2$H$^+$, C$_2$H, HC$_3$N, HN$^{13}$C.}
\renewcommand{\arraystretch}{1.2}
\tablehead{
 RMS   & N$_2$H$^+$ & N$_2$H$^+$    & C$_2$H      & HC$_3$N     & HN$^{13}$C & HNC\\
  name    & Group1  & Group2       & F = 2 - 1   &  J = 10 - 9  & J  = 1 - 0 & J  = 1 - 0\\
            & (K km s$^{-1}$) & (K km s$^{-1}$) & (K km s$^{-1}$) & (K km
 s$^{-1}$) & (K km s$^{-1}$) & (K km s$^{-1}$)
}\startdata
  & & & HII regions\\
  \hline\\
G300.5047-00.1745 &...         &...           & 2.53 (0.26 ) & ...            & ...           &4.43  (0.22)         \\

G301.1364-00.2249 &1.23(0.21)  &3.22 (0.20 )  & 6.31 (0.18 ) & 4.37 (0.17 )   & ...           &12.23 (0.17)         \\

G301.8147+00.7808 &...         &1.67 (0.14 )  & 1.52 (0.16 ) & ...            & ...           &3.26  (0.13)         \\
G302.0319-00.0613 &...         &...           & 3.37 (0.36 ) & ...            & ...           &4.19  (0.32)         \\
G302.4867-00.0308 &...         &...           & ...          & ...            & ...           &3.70  (0.23)         \\

G307.5606-00.5871 &1.27(0.20)  &7.87 (0.34 )  & 2.14 (0.21 ) & 2.08 (0.19 )   & ...           &9.33  (0.24)         \\
G307.6138-00.2559 &1.51(0.23)  &3.15 (0.20 )  & 4.83 (0.20 ) & ...            & ...           &7.33  (0.17)         \\

G309.9206+00.4790 &...         &...           & ...          & ...            & ...           &9.83  (0.66)          \\
G310.1420+00.7583 &1.04(0.13)  &4.27 (0.17 )  & 2.73 (0.18 ) & 2.57 (0.21 )   & ...           &9.28  (0.20)          \\
G311.6264+00.2897 &...         &...           & ...          &...             & ...           &...                  \\
G316.1386-00.5009 &1.23(0.13)  &5.45 (0.15 )  & 3.67 (0.17 ) & 1.41 (0.13 )   & ...           &8.01  (0.18)          \\
G317.4298-00.5612 &...         &...           & ...          & ...            & ...           &1.60  (0.17)          \\
G318.9148-00.1647 &...         &...           & ...          & 0.97 (0.16 )   & ...           &7.73  (0.54)          \\

G320.7779+00.2412 &...         &1.52 (0.19 )  & ...          & ...            & ...           &1.51  (0.19)          \\
G321.7209+01.1711 &1.47(0.16)  &5.78 (0.21 )  & 3.56 (0.20 ) & 1.64 (0.19 )   & ...           &7.74  (1.81)          \\
G323.4468+00.0968 &1.15(0.20)  &4.78 (0.17 )  & ...          & ...            & ...           &5.92  (0.20)          \\

G324.1997+00.1192 &...         &...           & 2.82 (0.20 ) & ...            & ...           &10.41 (0.25)            \\
G326.4477-00.7485 &1.69(0.19)  &6.79 (0.18 )  & ...          & 1.77 (0.17 )   & ...           &4.55  (0.21)            \\
G326.4719-00.3777 &1.74(0.17)  &6.64 (0.20 )  & 4.41 (0.22 ) & 1.97 (0.15 )   & ...           &8.90  (0.18)            \\
G326.7249+00.6159 &2.02(0.18)  &9.74 (0.21 )  & 4.9  (0.25 ) & 3.30 (0.22 )   & ...           &17.34 (0.24)            \\
G328.3067+00.4308 &...         &...           & 6.15 (0.24 ) & 2.09 (0.19 )   & ...           &18.22 (0.88)           \\
G328.8074+00.6324 &1.80(0.15)  &7.32 (0.17 )  & 4.85 (0.17 ) & 7.92 (0.21 )   & 2.43 (0.18 )  &14.92 (0.19)         \\
G328.9580+00.5671 &0.96(0.13)  &5.09 (0.19 )  & 4.17 (0.21 ) & 1.82 (0.15 )   & ...           &6.71  (0.18)          \\
G329.3371+00.1469 &0.95(0.18)  &5.52 (0.26 )  & 5.09 (0.22 ) & 3.4  (0.16 )   & 1.13 (0.16 )  &20.92 (0.25)         \\
G329.4055-00.4574 &1.98(0.18)  &6.27 (0.57 )  & 4.51 (0.19 ) & 4.12 (0.16 )   & 1.41 (0.21 )  &8.64  (0.36)         \\
G329.4211-00.1631 &1.80(0.15)  &8.01 (0.17 )  & 2.48 (0.18 ) & 1.37 (0.15 )   & ...           &9.65  (0.17)        \\
G329.4761+00.8414 &0.50(0.15)  &1.78 (0.18 )  & ...          & ...            & ...           &4.20  (0.19)        \\
G329.8145+00.1411 &1.55(0.22)  &6.19 (0.29 )  & ...          & ...            & ...           &5.96  (0.26)        \\
G330.2845+00.4933 &0.52(0.09)  &3.26 (0.18 )  & ...          & ...            & ...           &4.88  (0.21)        \\
G330.2935-00.3946 &...         &...           & 2.76 (0.19 ) & ...            & ...           &9.72  (0.21)          \\
G330.8845-00.3721 &2.09(0.23)  &7.21 (0.24 )  & 8.24 (0.22 ) & 6.4  (0.21 )   & 1.76 (0.19 )  &18.61 (0.22)         \\
G330.9544-00.1817 &...         &...           & 6.11 (0.24 ) & 8.11 (0.26 )   & 2.52 (0.25 )  &24.85 (2.10)         \\
G331.1282-00.2436 &2.45(0.20)  &12.57(0.32 )  & 4.07 (0.23 ) & 4.75 (0.21 )   & ...           &13.68 (0.93)        \\
G331.4181-00.3546 &2.29(0.33)  &7.78 (0.35 )  & ...          & ...            & ...           &6.99  (0.38)        \\
G331.5582-00.1206 &1.81(0.15)  &7.59 (0.18 )  & ...          & 2.27 (0.19 )   & ...           &5.58  (0.17)        \\
G332.1544-00.4487 &...         &...           & 5.85 (0.26 ) & ...            & ...           &10.40 (0.12)        \\
G332.2944-00.0962 &2.20(0.12)  &9.11 (0.15 )  & 4.25 (0.18 ) & 2.52 (0.16 )   & ...           &9.19  (0.18)        \\
G332.8256-00.5498 &3.52(0.11)  &17.28(1.06 )  & 10.37(0.28 ) & 9.80 (0.22 )   & 3.5  (0.23 )  &29.73 (0.25)         \\
G333.0299-00.0645 &0.49(0.10)  &2.42 (0.12 )  & 3.28 (0.16 ) & 1.62 (0.12 )   & ...           &4.25  (0.14)         \\
G333.0494+00.0324 &0.43(0.09)  &2.32 (0.11 )  & 2.30 (0.16 ) & ...            & ...           &4.97  (0.14)         \\
G333.1642-00.0994 &2.67(0.28)  &4.97 (0.30 )  & ...          & ...            & ...           &4.69  (0.61)         \\
G336.3684-00.0033 &...         & ...          & 3.16 (0.30 ) & ...            & ...           &16.38 (0.31)         \\
G337.0047+00.3226 &0.39(0.12)  &3.37 (0.22 )  & ...          & ...            & ...           &4.81  (0.18)         \\
G337.4050-00.4071 &2.51(0.17)  &10.53(0.17 )  & 6.75 (0.19 ) & 7.31 (0.17 )   & 2.08 (0.17 )  &21.05 (0.18)         \\
G337.8442-00.3748 &0.78(0.15)  &5.16 (0.15 )  & 2.74 (0.19 ) & 1.69 (0.12 )   & ...           &7.28  (0.14)         \\
G337.9947+00.0770 &...         &...           & 3.02 (0.17 ) & ...            & ...           &9.27  (0.17)         \\
G338.0715+00.0126 &...         &...           & ...          & ...            & ...           &...                 \\

G339.1052+00.1490 &0.94(0.10)  &5.02 (0.14 )  &...           & ...            & ...           &5.99  (0.15)         \\
G340.0708+00.9267 &1.37(0.16)  &6.21 (0.23 )  & 3.16 (0.18 ) & 1.75 (0.15 )   & ...           &8.97  (0.17)         \\
G340.2480-00.3725 &5.02(0.13)  &23.22(0.18 )  & 4.31 (0.20 ) & 5.70 (0.16 )   & 3.11 (0.16 )  &18.87 (0.82)         \\
G342.7057+00.1260 &3.45(0.19)  &16.00(0.29 )  & 6.42 (0.22 ) & 5.84 (0.18 )   & 2.05 (0.17 )  &15.52 (0.57)         \\
G344.4257+00.0451 &1.56(0.13)  &7.08 (0.16 )  & 3.94 (0.23 ) & 2.54 (0.13 )   & 1.51 (0.16 )  &9.48  (0.18)         \\
G345.0034-00.2240 &3.89(0.16)  &11.77(0.42 )  & 3.54 (0.23 ) & 10.46(0.27 )   & 3.23 (0.23 )  &12.33 (0.56)         \\
G345.4881+00.3148 &3.27(0.24)  &15.17(0.53 )  & 11.53(0.30 ) & 9.27 (0.31 )   & 3.17 (0.28 )  &35.97 (0.56)         \\
G345.6495+00.0084 &...         & ...          &...           & ...            & ...           &5.19  (0.32)         \\
G346.0774-00.0562 &1.88(0.21)  &6.85 (0.26 )  &...           & ...            & ...           &5.05  (0.27)         \\
G346.5235+00.0839 &...         &...           &...           & ...            & ...           &3.48  (0.18)         \\
G347.8707+00.0146 &...         &...           & 2.95 (0.20 ) & ...            & ...           &9.48  (0.24)         \\
G348.8922-00.1787 &...         &2.52 (0.16 )  & 1.91 (0.19 ) & ...            & ...           &6.36  (0.19)         \\
  \hline\\
& & & MYSOs\\
  \hline\\
G305.2017+00.2072 &...          &...           &6.90 (0.26)   &7.74  (0.27)    &...           &21.23 (0.27)        \\
G310.0135+00.3892 &1.56(0.13)   &5.32 (0.14)   &2.81 (0.14)   &1.73  (0.13)    &...           &6.02  (0.13)        \\
G313.7654-00.8620 &3.18(0.16)   &15.75(0.22)   &4.07 (0.24)   &3.86  (0.22)    &1.66  (0.17)  &16.38 (0.21)        \\
G314.3197+00.1125 &1.45(0.15)   &7.28 (0.20)   &...           &1.20  (0.17)    &...           &3.76  (0.27)        \\
G318.9480-00.1969 &3.06(0.13)   &13.59(0.15)   &3.25 (0.18)   &4.19  (0.17)    &1.78  (0.19)  &17.42 (2.38)        \\
G320.2437-00.5619 &0.60(0.13)   &2.92 (0.18)   &...           &...             &...           &2.01  (0.17)        \\
G326.4755+00.6947 &4.08(0.21)   &17.31(0.25)   &5.06 (0.28)   &4.81  (0.24)    &1.66  (0.19)  &19.31 (1.48)        \\
G326.5437+00.1684 &1.30(0.21)   &7.13 (0.30)   &1.83 (1.26)   &...             &...           &7.59  (0.30)        \\
G327.1192+00.5103 &0.64(0.14)   &3.69 (0.19)   &2.03 (0.26)   &...             &...           &3.04  (0.26)       \\
G329.0663-00.3081 &2.72(0.16)   &13.58(0.29)   &2.65 (0.21)   &4.23  (0.21)    &2.61  (0.21)  &11.10 (0.21)       \\
G330.9288-00.4070 &1.44(0.11)   &5.72 (0.14)   &1.53 (0.14)   &1.20  (0.16)    &0.95  (0.14)  &5.95  (0.14)      \\
G332.4683-00.5228 &1.88(0.13)   &11.23(0.17)   &4.16 (0.21)   &3.76  (0.19)    &1.66  (0.17)  &13.63 (0.25)      \\
G332.9636-00.6800 &4.12(0.32)   &12.80(0.66)   &5.69 (0.32)   &4.87  (0.32)    &...           &14.55 (0.46)      \\
G333.3151+00.1053 &2.68(0.17)   &13.25(0.21)   &3.40 (0.30)   &3.20  (0.22)    &1.2   (0.23)  &14.48 (0.26)      \\
G333.3752-00.2015 &0.44(0.14)   &2.31 (0.26)   &...           &...             &...           &3.57  (0.26)      \\
G335.0611-00.4261 &2.12(0.26)   &9.05 (0.30)   &...           &...             &...           &7.46  (0.46)      \\
G337.1555-00.3951 &...          &...           &...           &...             &...           &...               \\
G338.2801+00.5419 &1.59(0.19)   &8.91 (0.22)   &...           &...             &...           &11.35 (0.53)      \\
G338.9196+00.5495 &8.87(0.26)   &40.16(0.46)   &6.73 (0.29)   &8.34  (0.26)    &4.19  (0.27)  &25.37 (0.87)      \\
G339.6221-00.1209 &1.69(0.14)   &9.72 (0.16)   &3.47 (0.30)   &2.41  (0.17)    &1.84  (0.24)  &12.65 (0.62)      \\
G341.1281-00.3466 &0.90(0.22)   &5.04 (0.23)   &...           & ...            &...           &7.94  (0.24)      \\
G341.2182-00.2136 &2.91(0.23)   &14.41(0.26)   &3.04 (0.11)   &2.97  (0.25)    &...           &9.42  (0.31)      \\
G342.9583-00.3180 &1.12(0.20)   &4.25 (0.25)   &...           &...             &...           &4.05  (0.25)      \\
G343.5213-00.5171 &1.98(0.12)   &9.38 (0.14)   &...           &1.18  (0.10)    &...           &6.32  (0.16)      \\
G345.2619-00.4188 &1.30(0.19)   &3.33 (0.16)   &1.45 (0.17)   &0.81  (0.13)    &...           &3.74  (0.20)      \\
G345.7172+00.8166 &2.99(0.13)   &11.18(0.16)   &2.84 (0.16)   &2.74  (0.14)    &1.6   (0.15)  &8.95  (0.36)      \\
G346.4809+00.1320 &...          &1.82 (0.19)   &...           &...             &...           &3.69  (0.62)      \\
G348.6491+00.0225 &1.70(0.18)   &2.94 (0.25)   &...           &...             &...           &6.95  (0.21)      \\
\enddata

\end{deluxetable}

\clearpage


\tabletypesize{\scriptsize}

\setlength{\tabcolsep}{0.1in}

\begin{deluxetable}{lrrrrrr}
\tablewidth{0pt} \tablecaption{ Peak intensities of N$_2$H$^+$,
C$_2$H, HC$_3$N, HN$^{13}$C.}
\renewcommand{\arraystretch}{1.2}
\tablehead{
 RMS   & N$_2$H$^+$ & C$_2$H    & C$_2$H      & HC$_3$N     & HN$^{13}$C & HNC\\
  name    & Group2  & F = 1 - 0       & F = 2 - 1   &  J = 10 - 9  & J  = 1 - 0 & J  = 1 - 0\\
            & (K ) & (K ) & (K ) & (K ) & (K ) & (K )
}\startdata
  & & & HII regions\\
  \hline\\
G300.5047-00.1745 &0.25   &0.21   & 0.47  &...     &...    &1.03          \\

G301.1364-00.2249 &0.79   &0.85   & 1.46  &0.96    &...    &2.75          \\

G301.8147+00.7808 &0.79   &0.25   & 0.61  &...     &...    &1.41          \\
G302.0319-00.0613 &...    &...    & 1.03  &...     &...    &1.46          \\
G302.4867-00.0308 &...    &...    &...    &...     &...    &0.98          \\

G307.5606-00.5871 &1.36   &0.25   & 0.39  &0.46    &...    &1.66          \\
G307.6138-00.2559 &0.89   &0.57   & 1.32  &...     &...    &2.00          \\

G309.9206+00.4790 &...    &...    &...    &...     &...    &1.56           \\
G310.1420+00.7583 &1.28   &0.38   & 0.93  &0.69    &...    &2.36           \\
G311.6264+00.2897 &...    &...    &...    &...     &...    &...           \\
G316.1386-00.5009 &1.68   &0.68   & 1.01  &0.55    &...    &1.83           \\
G317.4298-00.5612 &...    &...    &...    &...     &...    &0.46           \\
G318.9148-00.1647 &...    &...    &...    &0.30    &...    &1.03           \\

G320.7779+00.2412 &0.51   &...    &...    &...     &...    &0.87           \\
G321.7209+01.1711 &1.47   &0.67   & 1.23  &0.59    &...    &1.89           \\
G323.4468+00.0968 &1.91   &...    &...    &...     &...    &1.72           \\

G324.1997+00.1192 &...    &0.28   & 0.62  &...     &...    &1.70             \\
G326.4477-00.7485 &1.88   &...    &...    &0.46    &...    &0.93             \\
G326.4719-00.3777 &2.00   &0.54   & 1.06  &0.76    &...    &2.12             \\
G326.7249+00.6159 &2.70   &0.64   & 1.30  &0.82    &...    &4.05             \\
G328.3067+00.4308 &...    &0.61   & 0.94  &0.43    &...    &1.18            \\
G328.8074+00.6324 &2.24   &1.05   & 1.59  &1.91    &0.69   &4.69          \\
G328.9580+00.5671 &1.35   &0.50   & 0.90  &0.63    &...    &1.52           \\
G329.3371+00.1469 &1.45   &0.52   & 0.97  &0.83    &0.37   &4.01          \\
G329.4055-00.4574 &1.64   &0.48   & 0.90  &1.00    &0.32   &1.71          \\
G329.4211-00.1631 &2.72   &0.34   & 1.00  &0.62    &...    &2.72         \\
G329.4761+00.8414 &0.60   &...    &...    &...     &...    &1.09         \\
G329.8145+00.1411 &1.55   &...    &...    &...     &...    &1.55         \\
G330.2845+00.4933 &1.13   &...    &...    &...     &...    &1.30         \\
G330.2935-00.3946 &...    &0.38   & 0.63  &...     &...    &1.51           \\
G330.8845-00.3721 &1.60   &0.85   & 1.62  &1.25    &0.40   &2.90          \\
G330.9544-00.1817 &1.08   &0.51   & 1.00  &0.90    &0.38   &2.07          \\
G331.1282-00.2436 &2.11   &0.42   & 0.66  &0.79    &...    &1.33         \\
G331.4181-00.3546 &2.55   &...    &...    &...     &...    &2.01         \\
G331.5582-00.1206 &2.43   &...    &...    &0.55    &...    &1.87         \\
G332.1544-00.4487 &0.71   &0.39   & 0.78  &...     &...    &1.49         \\
G332.2944-00.0962 &2.85   &0.63   & 1.15  &0.77    &...    &2.20         \\
G332.8256-00.5498 &2.68   &0.74   & 1.30  &1.48    &0.58   &3.26          \\
G333.0299-00.0645 &0.85   &0.46   & 1.00  &0.65    &...    &1.13          \\
G333.0494+00.0324 &1.00   &0.43   & 0.80  &...     &...    &1.74          \\
G333.1642-00.0994 &1.61   &...    &...    &...     &...    &0.72          \\
G336.3684-00.0033 &...    &0.24   & 0.68  &...     &...    &2.54          \\
G337.0047+00.3226 &0.92   &...    &...    &...     &...    &1.17          \\
G337.4050-00.4071 &2.76   &0.86   & 1.65  &1.82    &0.54   &3.87          \\
G337.8442-00.3748 &2.00   &0.52   & 0.98  &1.00    &...    &2.29          \\
G337.9947+00.0770 &...    &0.50   & 0.83  &...     &...    &2.21          \\
G338.0715+00.0126 &1.03   &...    &...    &...     &...    &...          \\

G339.1052+00.1490 &1.59   &...    &...    &...     &...    &1.35          \\
G340.0708+00.9267 &1.73   &0.45   & 0.90  &0.52    &...    &2.40          \\
G340.2480-00.3725 &5.88   &0.67   & 1.10  &1.60    &0.96   &3.41          \\
G342.7057+00.1260 &3.01   &0.87   & 1.45  &1.74    &0.70   &3.55          \\
G344.4257+00.0451 &2.06   &0.55   & 0.92  &0.96    &0.47   &2.13          \\
G345.0034-00.2240 &2.64   &0.30   & 0.61  &1.36    &0.59   &1.09          \\
G345.4881+00.3148 &4.10   &1.15   & 2.29  &1.95    &0.65   &5.15          \\
G345.6495+00.0084 &...    &...    &...    &...     &...    &0.57          \\
G346.0774-00.0562 &1.99   &...    &...    &...     &...    &1.22          \\
G346.5235+00.0839 &...    &...    &...    &...     &...    &0.70          \\
G347.8707+00.0146 &0.92   &0.37   & 0.64  &...     &...    &1.34          \\
G348.8922-00.1787 &0.68   &0.22   & 0.44  &...     &...    &1.18          \\
  \hline\\
& & & MYSOs\\
  \hline\\
G305.2017+00.2072 &...     &0.41    &1.03    &0.96     &...    &2.86         \\
G310.0135+00.3892 &1.90    &0.52    &1.31    &0.88     &...    &2.78         \\
G313.7654-00.8620 &3.92    &0.49    &0.91    &0.90     &0.62   &3.42         \\
G314.3197+00.1125 &1.99    &...     &...     &0.42     &...    &0.75         \\
G318.9480-00.1969 &4.17    &0.53    &1.05    &1.24     &0.53   &2.72         \\
G320.2437-00.5619 &0.97    &...     &...     &...      &...    &0.59         \\
G326.4755+00.6947 &4.83    &0.49    &1.20    &1.23     &0.59   &2.68         \\
G326.5437+00.1684 &2.00    &0.43    &0.52    &...      &...    &1.83         \\
G327.1192+00.5103 &1.03    &0.34    &0.50    &...      &...    &0.69        \\
G329.0663-00.3081 &2.40    &0.38    &0.55    &0.73     &0.47   &2.26        \\
G330.9288-00.4070 &1.92    &0.52    &0.72    &0.43     &0.49   &1.91       \\
G332.4683-00.5228 &3.21    &0.62    &1.04    &1.04     &0.58   &2.45       \\
G332.9636-00.6800 &3.18    &0.44    &0.99    &1.29     &...    &1.78       \\
G333.3151+00.1053 &3.63    &0.54    &0.76    &0.94     &0.50   &3.13       \\
G333.3752-00.2015 &0.79    &...     &...     &...      &...    &0.94       \\
G335.0611-00.4261 &2.78    &...     &...     &...      &...    &1.60       \\
G337.1555-00.3951 &...     &...     &...     &...      &...    &...        \\
G338.2801+00.5419 &2.16    &...     &...     &...      &...    &2.71       \\
G338.9196+00.5495 &5.14    &0.53    &0.91    &1.25     &0.61   &3.41       \\
G339.6221-00.1209 &3.39    &0.43    &0.48    &0.84     &0.61   &2.52       \\
G341.1281-00.3466 &1.63    &...     &...     &...      &...    &2.33       \\
G341.2182-00.2136 &3.97    &0.44    &1.08    &0.88     &...    &2.33       \\
G342.9583-00.3180 &1.68    &...     &...     &...      &...    &1.77       \\
G343.5213-00.5171 &2.66    &...     &...     &0.60     &...    &1.31       \\
G345.2619-00.4188 &1.20    &...     &0.67    &0.51     &...    &1.20       \\
G345.7172+00.8166 &3.41    &0.81    &1.01    &1.22     &0.70   &2.57       \\
G346.4809+00.1320 &0.52    &...     &...     &...      &...    &0.77       \\
G348.6491+00.0225 &0.80    &...     &...     &...      &...    &0.67       \\
\enddata

\end{deluxetable}

\clearpage

\tabletypesize{\scriptsize}

\setlength{\tabcolsep}{0.1in}

\begin{deluxetable}{lrrrr}
\tablewidth{0pt} \tablecaption{ Column densities of N$_2$H$^+$,
C$_2$H, HC$_3$N, HNC.}
\renewcommand{\arraystretch}{1.2}
\tablehead{
RMS   & N$_2$H$^+$    & C$_2$H      & HC$_3$N     & HNC \\
name    & 10$^{13}$ cm$^{-2}$  & 10$^{14}$ cm$^{-2}$ & 10$^{13}$ cm$^{-2}$ & 10$^{14}$ cm$^{-2}$\\
}\startdata

  & & HII regions\\
    \hline\\
G300.5047-00.1745  &...         & 4.37 (0.55) &...          &...                    \\

G301.1364-00.2249  &0.73(0.06)  & 10.51(0.51) &1.77(0.07)   &...              \\

G301.8147+00.7808  &0.38(0.06)  & 1.81 (0.36) &...          &...              \\
G302.0319-00.0613  &...         & 4.57 (0.61) &...          &...               \\
G302.4867-00.0308  &...         &...          &...          &...               \\

G307.5606-00.5871  &1.68(0.08)  & 4.07 (0.49) &0.84(0.07)   &...              \\
G307.6138-00.2559  &0.71(0.08)  & 5.76 (0.40) &...          &...              \\

G309.9206+00.4790  &...         &...          &...          &...              \\
G310.1420+00.7583  &0.96(0.09)  & 3.26 (0.31) &1.04(0.08)   &...              \\
G311.6264+00.2897  &...         &...          &...          &...              \\
G316.1386-00.5009  &1.16(0.07)  & 7.81 (0.59) &0.57(0.05)   &...              \\
G317.4298-00.5612  &...         &...          &...          &...              \\
G318.9148-00.1647  &...         &...          &0.4 (0.08)   &...              \\

G320.7779+00.2412  &0.31(0.05)  &...          &...          &...              \\
G321.7209+01.1711  &1.23(0.16)  & 4.67 (0.67) &0.66(0.08)   &...              \\
G323.4468+00.0968  &0.93(0.05)  &...          &...          &...             \\

G324.1997+00.1192  &...         & 4.01 (0.48) &...          &...              \\
G326.4477-00.7485  &1.32(0.08)  &...          &0.71(0.07)   &...              \\
G326.4719-00.3777  &1.46(0.08)  & 5.49 (0.41) &0.79(0.06)   &...               \\
G326.7249+00.6159  &2.32(0.11)  & 6.16  0.48  &1.33(0.11)   &...               \\
G328.3067+00.4308  &...         & 16.11(0.98) &0.87(0.13)   &...               \\
G328.8074+00.6324  &1.47(0.13)  & 9.15 (0.85) &3.21(0.06)   &2.83   (0.17 )          \\
G328.9580+00.5671  &1.05(0.07)  & 5.65 (0.56) &0.73(0.06)   &...                     \\
G329.3371+00.1469  &1.56(0.1 )  & 9.00 (0.60) &1.39(0.15)   &3.28   (0.11 )          \\
G329.4055-00.4574  &1.34(0.2 )  & 6.03 (0.61) &1.65(0.08)   &2.25   (0.20 )          \\
G329.4211-00.1631  &1.61(0.13)  & 2.64 (0.36) &0.55(0.06)   &...                    \\
G329.4761+00.8414  &0.41(0.05)  &...          &...          &...                    \\
G329.8145+00.1411  &1.17(0.07)  &...          &...          &...                    \\
G330.2845+00.4933  &0.66(0.03)  &...          &...          &...                    \\
G330.2935-00.3946  &...         & 3.54 (0.34) &...          &...                    \\
G330.8845-00.3721  &1.58(0.1 )  & 10.6 (0.52) &2.62(0.05)   &3.54   (0.13 )          \\
G330.9544-00.1817  &...         & 9.19 (0.58) &3.31(0.29)   &7.60   (0.80 )          \\
G331.1282-00.2436  &2.48(0.11)  & 7.04 (0.53) &1.93(0.06)   &...                   \\
G331.4181-00.3546  &1.47(0.31)  &...          &...          &...                   \\
G331.5582-00.1206  &1.58(0.07)  &...          &0.92(0.06)   &...                   \\
G332.1544-00.4487  &...         & 8.54 (0.54) &...          &...                   \\
G332.2944-00.0962  &1.78(0.13)  & 5.43 (0.57) &1.02(0.05)   &...                   \\
G332.8256-00.5498  &4.01(0.59)  & 16.66(1.81) &3.98(0.16)   &7.89   (0.58 )          \\
G333.0299-00.0645  &0.55(0.07)  & 3.91 (0.53) &0.66(0.06)   &...                    \\
G333.0494+00.0324  &0.50(0.04)  & 3.06 (0.40) &...          &...                    \\
G333.1642-00.0994  &1.06(0.13)  &...          &...          &...                    \\
G336.3684-00.0033  &...         & 3.16 (0.41) &...          &...                    \\
G337.0047+00.3226  &0.7 (0.11)  &...          &...          &...                    \\
G337.4050-00.4071  &2.28(0.08)  & 8.57 (0.41) &2.94(0.07)   &3.97   (0.07 )          \\
G337.8442-00.3748  &1.19(0.07)  & 3.91 (0.37) &0.68(0.05)   &...                 \\
G337.9947+00.0770  &...         & 4.68 (0.37) &...          &...                 \\
G338.0715+00.0126  &...         &...          &...          &...                 \\

G339.1052+00.1490  &1.01(0.06)  &...          &...          &...                 \\
G340.0708+00.9267  &1.29(0.12)  & 3.47 (0.40) &0.7 (0.07)   &...                 \\
G340.2480-00.3725  &4.24(0.17)  & 6.36 (0.51) &2.39(0.04)   &6.90   (0.47 )          \\
G342.7057+00.1260  &3.24(0.12)  & 9.98 (0.56) &2.43(0.04)   &4.07   (0.22 )          \\
G344.4257+00.0451  &1.68(0.07)  & 7.39 (0.59) &1.04(0.06)   &2.94   (0.11 )          \\
G345.0034-00.2240  &2.42(0.93)  & 3.86 (1.59) &4.25(0.32)   &11.48  (3.64 )          \\
G345.4881+00.3148  &3.33(0.21)  & 13.39(0.72) &3.72(0.16)   &4.62   (0.21 )          \\
G345.6495+00.0084  &...         &...          &...          &...                     \\
G346.0774-00.0562  &1.34(0.22)  &...          &...          &...                     \\
G346.5235+00.0839  &...         &...          &...          &...                     \\
G347.8707+00.0146  &...         & 4.44 (0.43) &...          &...                     \\
G348.8922-00.1787  &0.58(0.07)  & 2.34 (0.37) &...          &...                     \\
  \hline\\
& & MYSOs\\
  \hline\\
G305.2017+00.2072  &...            &7.74 (0.48)   &3.11 (0.11)    &...         \\
G310.0135+00.3892  &1.91 (0.10 )   &3.26 (0.26)   &0.69 (0.06)    &...         \\
G313.7654-00.8620  &3.00 (0.24 )   &4.61 (0.58)   &1.55 (0.08)    &3.71 (0.23) \\
G314.3197+00.1125  &1.34 (0.08 )   &...           &0.49 (0.06)    &...         \\
G318.9480-00.1969  &3.22 (0.08 )   &3.65 (0.24)   &1.68 (0.08)    &4.61 (0.68) \\
G320.2437-00.5619  &0.47 (0.07 )   &...           &...            &...         \\
G326.4755+00.6947  &3.21 (0.11 )   &4.96 (0.37)   &1.95 (0.09)    &...         \\
G326.5437+00.1684  &1.18 (0.23 )   &5.63 (5.22)   &...            &...         \\
G327.1192+00.5103  &0.79 (0.06 )   &4.42 (0.71)   &...            &...         \\
G329.0663-00.3081  &2.20 (0.12 )   &4.66 (0.47)   &1.81 (0.07)    &2.61 (0.09) \\
G330.9288-00.4070  &1.32 (0.05 )   &3.15 (0.34)   &0.49 (0.07)    &1.91 (0.07) \\
G332.4683-00.5228  &1.99 (0.10 )   &5.79 (0.49)   &1.54 (0.06)    &3.98 (0.17) \\
G332.9636-00.6800  &4.30 (0.37 )   &5.52 (0.48)   &1.98 (0.18)    &...         \\
G333.3151+00.1053  &2.54 (0.12 )   &7.31 (0.89)   &1.29 (0.08)    &2.79 (0.12) \\
G333.3752-00.2015  &0.44 (0.08 )   &...           &...            &...         \\
G335.0611-00.4261  &2.05 (0.13 )   &...           &...            &...         \\
G337.1555-00.3951  &...            &...           &...            &...         \\
G338.2801+00.5419  &1.57 (0.08 )   &              &...            &...         \\
G338.9196+00.5495  &7.74 (0.36 )   &8.45 (0.67)   &3.46 (0.07)    &5.37 (0.29) \\
G339.6221-00.1209  &2.00 (0.09 )   &15.04(1.78)   &0.97 (0.07)    &4.26 (0.31) \\
G341.1281-00.3466  &0.96 (0.05 )   &...           &...            &...         \\
G341.2182-00.2136  &2.81 (0.05 )   &3.11 (0.16)   &1.19 (0.11)    &...         \\
G342.9583-00.3180  &1.08 (0.07 )   &...           &...            &...         \\
G343.5213-00.5171  &1.71 (0.14 )   &...           &0.49 (0.03)    &...         \\
G345.2619-00.4188  &1.37 (0.07 )   &1.31 (0.18)   &0.34 (0.05)    &...         \\
G345.7172+00.8166  &2.76 (0.13 )   &7.69 (0.71)   &1.14 (0.04)    &3.03 (0.19) \\
G346.4809+00.1320  &0.34 (0.16 )   &...           &...            &...        \\
G348.6491+00.0225  &2.26 (0.26 )   &...           &...            &...        \\
\enddata

\end{deluxetable}

\clearpage


\tabletypesize{\scriptsize}

\setlength{\tabcolsep}{0.1in}

\begin{deluxetable}{lrrrr}
\tablewidth{0pt} \tablecaption{Abundances of N$_2$H$^+$, C$_2$H,
HC$_3$N, HNC.}
\renewcommand{\arraystretch}{1.2}
\tablehead{
 RMS   & N$_2$H$^+$    & C$_2$H      & HC$_3$N     & HNC     \\
 name    &  10$^{-9}$ & 10$^{-8}$  & 10$^{-10}$ &  10$^{-8}$
}\startdata

 & & HII regions\\
    \hline\\
G300.5047-00.1745  &...           &0.56  (0.10  ) &...             &...                 \\

G301.1364-00.2249  &0.04 (0.01 )  &0.24  (0.03  ) &0.40  (0.05 )   &...                     \\

G301.8147+00.7808  &0.12 (0.04 )  &0.58  (0.21  ) &...             &...                     \\
G302.0319-00.0613  &...           &1.01  (0.22  ) &...             &...                     \\
G302.4867-00.0308  &...           &...            &...             &...                     \\

G307.5606-00.5871  &0.13 (0.01 )  &0.31  (0.05  ) &0.63  (0.09 )   &...                     \\
G307.6138-00.2559  &0.40 (0.08 )  &1.01  (0.15  ) &...             &...                     \\

G309.9206+00.4790  &...           &...            &...             &...                    \\
G310.1420+00.7583  &0.12 (0.02 )  &0.32  (0.06  ) &1.01  (0.16 )   &...                    \\
G311.6264+00.2897  &...           &...            &...             &...                    \\
G316.1386-00.5009  &0.18 (0.03 )  &1.07  (0.17  ) &0.78  (0.13 )   &...                    \\
G317.4298-00.5612  &...           &...            &...             &...                    \\
G318.9148-00.1647  &...           &...            &0.60  (0.22 )   &...                    \\

G320.7779+00.2412  &0.09 (0.03 )  &...            &...             &...                     \\
G321.7209+01.1711  &0.15 (0.04 )  &0.42  (0.13  ) &0.60  (0.16 )   &...                     \\
G323.4468+00.0968  &0.24 (0.03 )  &...            &...             &...                    \\

G324.1997+00.1192  &...           &0.26  (0.06  ) &...             &...                     \\
G326.4477-00.7485  &0.23 (0.03 )  &...            &0.93  (0.17 )   &...                     \\
G326.4719-00.3777  &0.19 (0.03 )  &0.51  (0.08  ) &0.73  (0.11 )   &...                      \\
G326.7249+00.6159  &0.15 (0.02 )  &0.39  (0.06  ) &0.85  (0.14 )   &...                      \\
G328.3067+00.4308  &...           &1.21  (0.17  ) &0.65  (0.15 )   &...                       \\
G328.8074+00.6324  &0.03 (0.01 )  &0.16  (0.03  ) &0.55  (0.07 )   &0.05 (0.01 )          \\
G328.9580+00.5671  &0.15 (0.02 )  &0.79  (0.14  ) &1.02  (0.17 )   &...                      \\
G329.3371+00.1469  &0.13 (0.02 )  &0.72  (0.11  ) &1.12  (0.21 )   &0.26 (0.03 )          \\
G329.4055-00.4574  &0.16 (0.05 )  &0.40  (0.10  ) &1.10  (0.21 )   &0.15 (0.04 )          \\
G329.4211-00.1631  &0.28 (0.05 )  &0.40  (0.10  ) &0.82  (0.18 )   &...                \\
G329.4761+00.8414  &0.21 (0.04 )  &...            &...             &...                \\
G329.8145+00.1411  &0.23 (0.03 )  &...            &...             &...                 \\
G330.2845+00.4933  &0.14 (0.02 )  &...            &...             &...                 \\
G330.2935-00.3946  &...           &0.32  (0.06  ) &...             &...                  \\
G330.8845-00.3721  &0.06 (0.01 )  &0.27  (0.03  ) &0.67  (0.06 )   &0.09 (0.01 )          \\
G330.9544-00.1817  &...           &0.14  (0.02  ) &0.50  (0.08 )   &0.11 (0.02 )          \\
G331.1282-00.2436  &0.08 (0.01 )  &0.22  (0.03  ) &0.60  (0.05 )   &...                  \\
G331.4181-00.3546  &0.26 (0.09 )  &...            &...             &...                  \\
G331.5582-00.1206  &0.10 (0.01 )  &...            &0.49  (0.07 )   &...                  \\
G332.1544-00.4487  &...           &1.14  (0.13  ) &...             &...                  \\
G332.2944-00.0962  &0.13 (0.02 )  &0.31  (0.07  ) &0.58  (0.09 )   &...                   \\
G332.8256-00.5498  &0.07 (0.02 )  &0.27  (0.07  ) &0.65  (0.12 )   &0.13 (0.03 )          \\
G333.0299-00.0645  &0.12 (0.04 )  &0.85  (0.27  ) &1.43  (0.38 )   &...                   \\
G333.0494+00.0324  &0.13 (0.02 )  &0.80  (0.19  ) &...             &...                   \\
G333.1642-00.0994  &0.81 (0.22 )  &...            &...             &...                   \\
G336.3684-00.0033  &...           &0.20  (0.04  ) &...             &...                   \\
G337.0047+00.3226  &0.14 (0.04 )  &...            &...             &...                    \\
G337.4050-00.4071  &0.07 (0.01 )  &0.22  (0.02  ) &0.77  (0.06 )   &0.10 (0.01 )          \\
G337.8442-00.3748  &0.19 (0.03 )  &0.63  (0.11  ) &1.09  (0.17 )   &...                   \\
G337.9947+00.0770  &...           &0.80  (0.13  ) &...             &...                   \\
G338.0715+00.0126  &...           &...            &...             &...                   \\

G339.1052+00.1490  &0.21 (0.03 )  &...            &...             &...                    \\
G340.0708+00.9267  &0.24 (0.05 )  &0.58  (0.13  ) &1.18  (0.24 )   &...                    \\
G340.2480-00.3725  &0.33 (0.04 )  &0.45  (0.07  ) &1.67  (0.16 )   &0.48 (0.07 )          \\
G342.7057+00.1260  &0.23 (0.03 )  &0.65  (0.09  ) &1.59  (0.15 )   &0.27 (0.04 )          \\
G344.4257+00.0451  &0.22 (0.03 )  &0.86  (0.14  ) &1.21  (0.17 )   &0.34 (0.04 )          \\
G345.0034-00.2240  &0.11 (0.09 )  &0.09  (0.08  ) &1.03  (0.45 )   &0.28 (0.21 )          \\
G345.4881+00.3148  &0.10 (0.01 )  &0.37  (0.05  ) &1.04  (0.13 )   &0.13 (0.02 )          \\
G345.6495+00.0084  &...           &...            &...             &...                    \\
G346.0774-00.0562  &0.33 (0.12 )  &...            &...             &...                    \\
G346.5235+00.0839  &...           &...            &...             &...                    \\
G347.8707+00.0146  &...           &0.61  (0.12  ) &...             &...                    \\
G348.8922-00.1787  &0.10 (0.02 )  &0.42  (0.11  ) &...             &...                     \\
  \hline\\
& & MYSOs\\
  \hline\\
G305.2017+00.2072  &...            &0.24 (0.03 )   &0.97  (0.11  )    &...               \\
G310.0135+00.3892  &0.23 (0.03 )   &0.40 (0.06 )   &0.84  (0.14  )    &...               \\
G313.7654-00.8620  &0.25 (0.05 )   &0.39 (0.09 )   &1.31  (0.20  )    &0.31 (0.05 )  \\
G314.3197+00.1125  &0.19 (0.03 )   &...            &0.68  (0.14  )    &...             \\
G318.9480-00.1969  &0.19 (0.02 )   &0.21 (0.03 )   &0.98  (0.12  )    &0.27 (0.06 )  \\
G320.2437-00.5619  &0.08 (0.02 )   &...            &...               &...             \\
G326.4755+00.6947  &0.08 (0.01 )   &0.13 (0.02 )   &0.51  (0.06  )    &...             \\
G326.5437+00.1684  &0.12 (0.05 )   &0.56 (0.73 )   &...               &...             \\
G327.1192+00.5103  &0.10 (0.02 )   &0.56 (0.14 )   &...               &...             \\
G329.0663-00.3081  &0.17 (0.02 )   &0.36 (0.07 )   &1.39  (0.16  )    &0.20 (0.02 )  \\
G330.9288-00.4070  &0.14 (0.02 )   &0.34 (0.07 )   &0.53  (0.12  )    &0.21 (0.02 )  \\
G332.4683-00.5228  &0.22 (0.03 )   &0.63 (0.10 )   &1.67  (0.20  )    &0.43 (0.05 )  \\
G332.9636-00.6800  &0.20 (0.03 )   &0.26 (0.04 )   &0.92  (0.16  )    &                \\
G333.3151+00.1053  &0.27 (0.04 )   &0.78 (0.18 )   &1.37  (0.23  )    &0.30 (0.04 )  \\
G333.3752-00.2015  &0.09 (0.03 )   &...            &...               &...             \\
G335.0611-00.4261  &0.14 (0.02 )   &...            &...               &...             \\
G337.1555-00.3951  &...            &...            &...               &...             \\
G338.2801+00.5419  &0.22 (0.03 )   &...            &...               &...             \\
G338.9196+00.5495  &0.13 (0.02 )   &0.15 (0.02 )   &0.59  (0.06  )    &0.09 (0.01 )  \\
G339.6221-00.1209  &0.26 (0.03 )   &1.96 (0.40 )   &1.27  (0.19  )    &0.56 (0.09 )  \\
G341.1281-00.3466  &0.14 (0.02 )   &,..            &...               &...             \\
G341.2182-00.2136  &0.22 (0.01 )   &0.24 (0.03 )   &0.94  (0.14  )    &...             \\
G342.9583-00.3180  &0.31 (0.04 )   &...            &...               &...             \\
G343.5213-00.5171  &0.26 (0.06 )   &...            &0.75  (0.15  )    &...             \\
G345.2619-00.4188  &0.30 (0.03 )   &0.28 (0.05 )   &0.74  (0.15  )    &...             \\
G345.7172+00.8166  &0.24 (0.03 )   &0.66 (0.12 )   &0.98  (0.11  )    &0.26 (0.04 )  \\
G346.4809+00.1320  &0.12 (0.08 )   &...            &...               &...             \\
G348.6491+00.0225B &0.46 (0.10 )   &...            &...               &...             \\
\enddata

\end{deluxetable}

\clearpage

\begin{table}
\begin{minipage}{20cm}
 \caption{\label{tab:test}Average column densities and molecular abundances compared with other works.}
 \centering
 \begin{tabular}{lrrccc}
  \hline
  \hline
 Different works & Type & N$_2$H$^+$ & C$_2$H & HC$_3$N & HNC \\
 \hline
  Column Densities (cm$^{-2}$) \\
 \hline
 This work & RMSs & 1.5 $\times$ 10$^{13}$ & 6.4 $\times$ 10$^{14}$ & 1.6 $\times$ 10$^{13}$ & 4.4 $\times$ 10$^{14}$ \\
 Sanhueza et al. (2012) & IRDCs & 1.6 $\times$ 10$^{13}$ & 2.4 $\times$ 10$^{14}$ & 4.8 $\times$ 10$^{13}$ & 2.4 $\times$ 10$^{14}$ \\
 Yu et al. (2015) & EGOs & 6.5 $\times$ 10$^{13}$ & 5.7 $\times$ 10$^{14}$ & ... & ... \\
 Chen et al. (2013) & EGOs & 3.0 $\times$ 10$^{13}$ & ... & 3.2 $\times$ 10$^{13}$ & ... \\
 \hline
 Molecular Abundances \\
 \hline
 This work & RMSs & 0.2 $\times$ 10$^{-9}$ & 0.5 $\times$ 10$^{-8}$ & 0.9 $\times$ 10$^{-10}$ & 0.2 $\times$ 10$^{-8}$  \\
 Vasyunina et al. (2011) & IRDCs & 2.8 $\times$ 10$^{-9}$ & 1.4 $\times$ 10$^{-8}$ & 5.4 $\times$ 10$^{-10}$ & 0.2 $\times$ 10$^{-8}$  \\
 Sanhueza et al. (2012) & IRDCs & 2.4 $\times$ 10$^{-9}$ & 3.7 $\times$ 10$^{-8}$ & 4.2 $\times$ 10$^{-10}$ & 3.7 $\times$ 10$^{-8}$  \\
 \hline
 \end{tabular}
 \label{tb:rotn}
\end{minipage}
\end{table}
\clearpage

\begin{figure}
\psfig{file=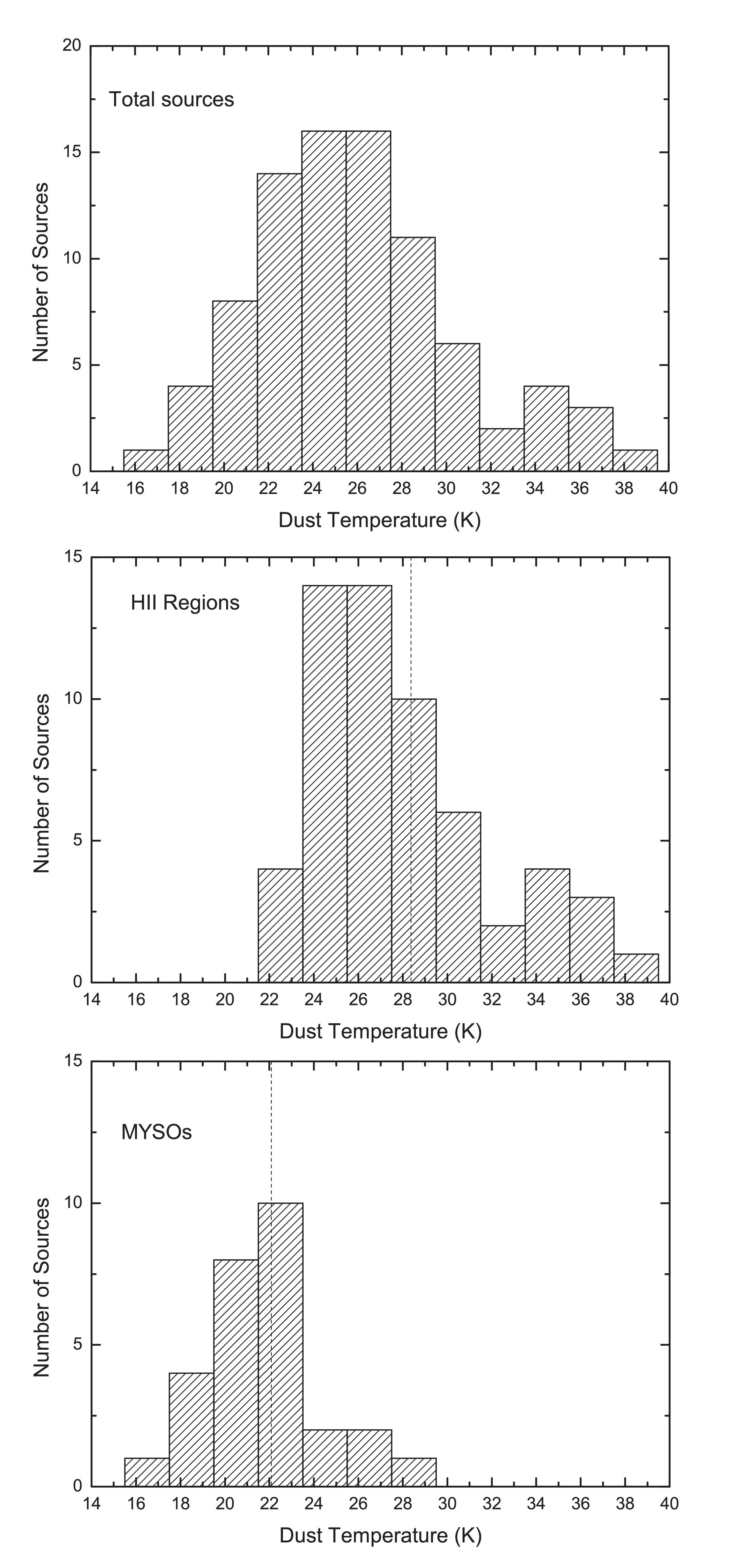,width=3in,height=7in} \caption{Dust
temperature distributions of total sources (top), HII regions
(middle) and MYSOs (bottom). The dashed vertical lines indicate
their mean temperatures.}
\end{figure}

\begin{figure}
\psfig{file=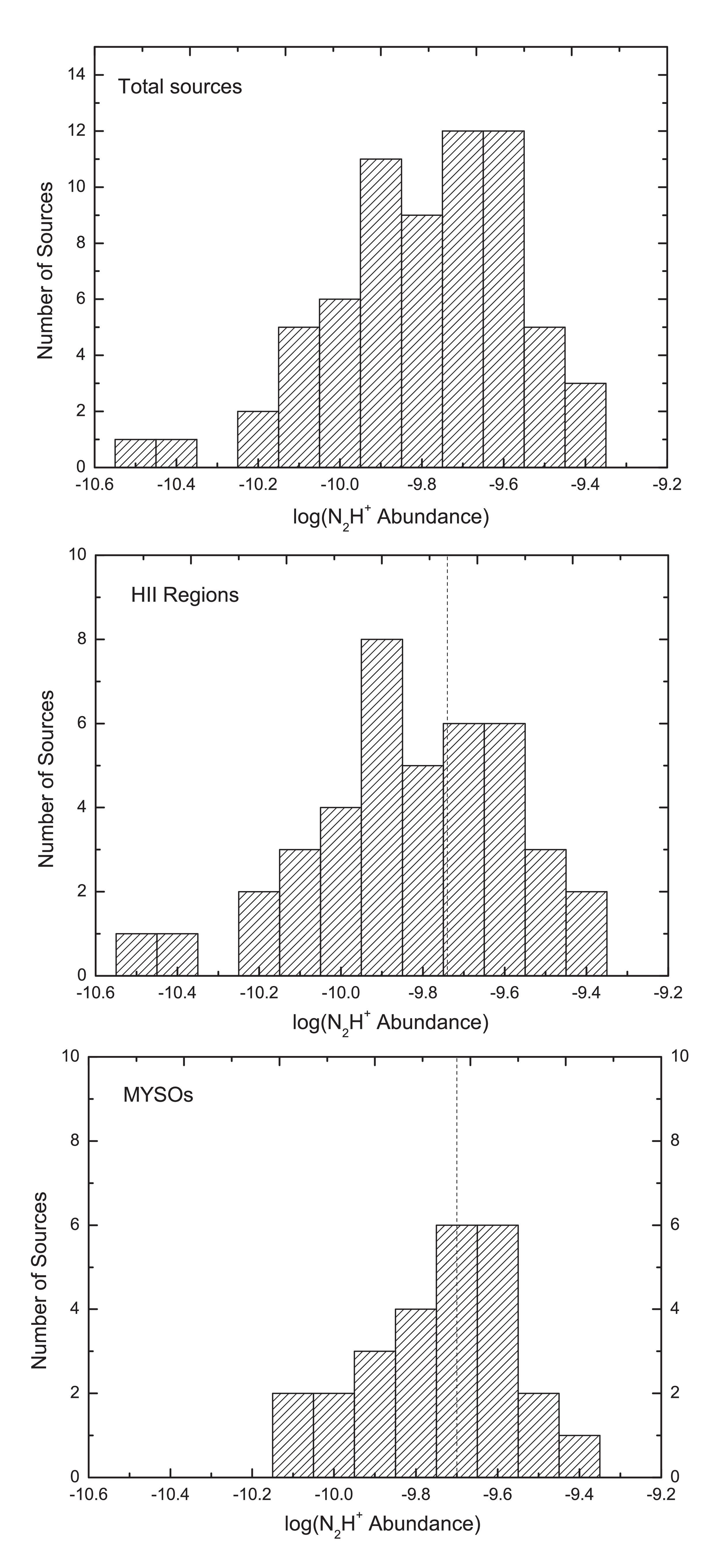,width=3in,height=7in}
 \caption{Histograms
of the number distributions of N$_2$H$^+$ abundances for our total
RMSs (top), MYSOs (middle) and HII regions (bottom). The dashed
lines indicate mean values.}
\end{figure}

\begin{figure}
\psfig{file=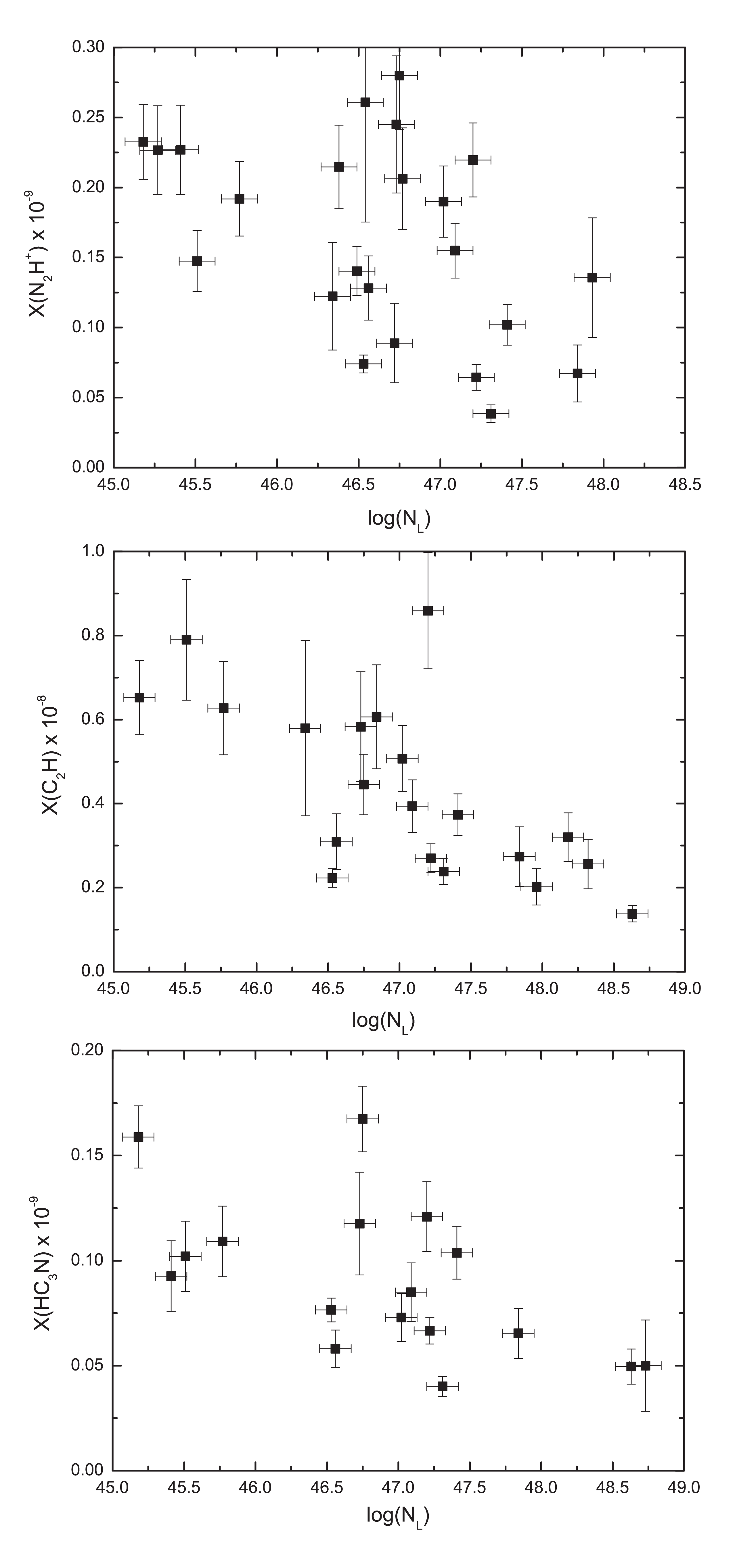,width=3in,height=7in}
 \caption{Fractional abundances of N$_2$H$^+$ (top), C$_2$H (middle) and HC$_3$N (bottom) plotted as a function of $N_L$ in logarithmic
scales.}
\end{figure}

\begin{figure}
\psfig{file=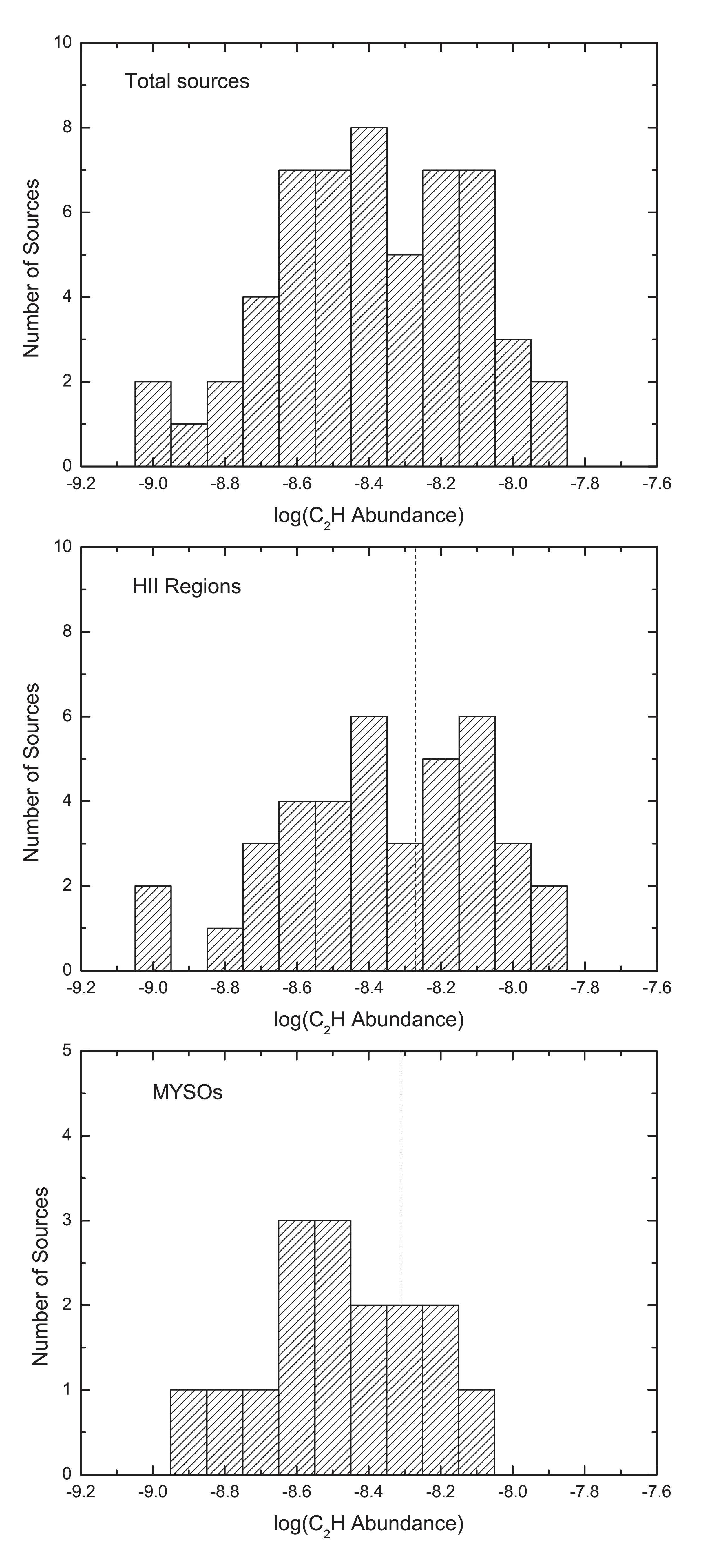,width=3.0in,height=7in}
 \caption{Histograms
of the number distributions of C$_2$H abundances for our total RMSs
(top), HII regions (middle) and MYSOs (bottom). The dashed lines
indicate mean values.}
\end{figure}

\begin{figure}
\psfig{file=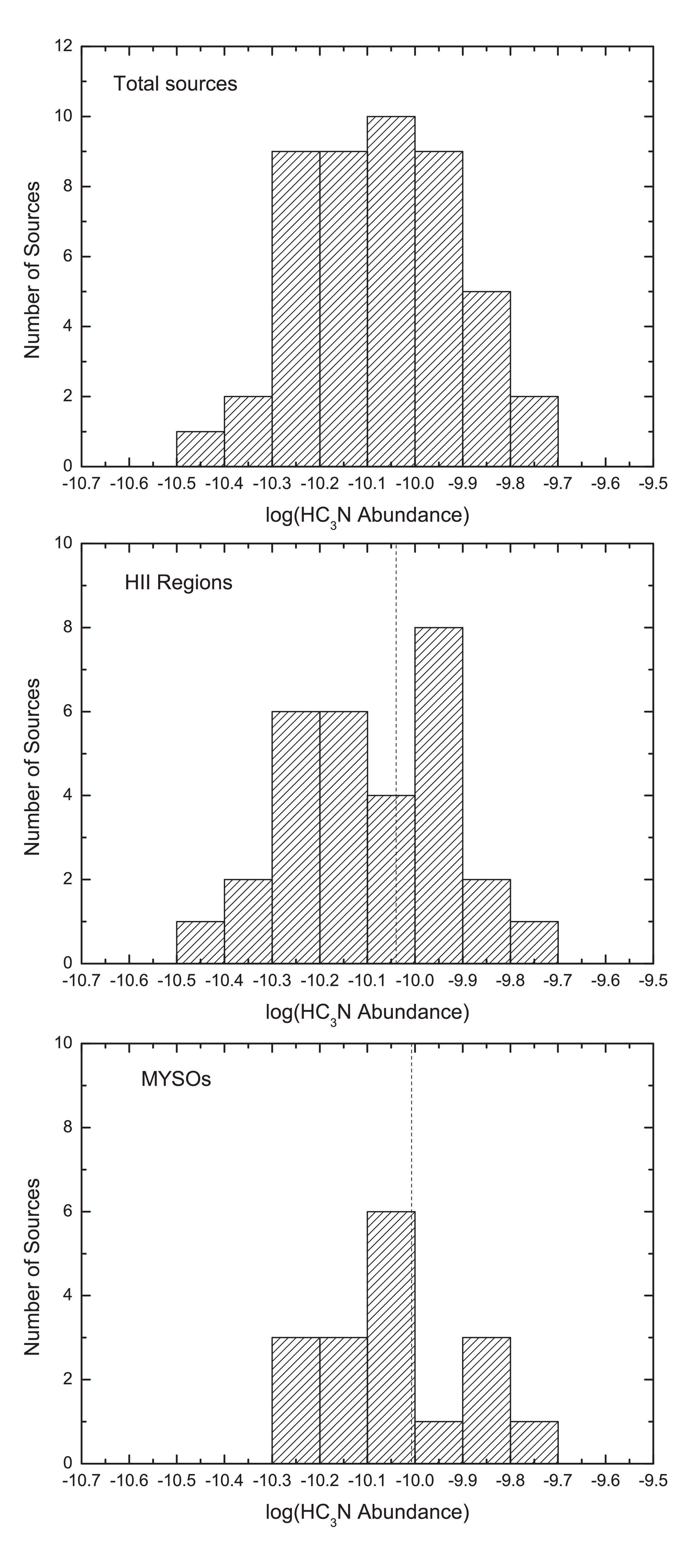,width=3.0in,height=7in}
 \caption{Histograms
of the number distributions of HC$_3$N abundances for our total RMSs
(top), HII regions (middle) and MYSOs (bottom). The dashed lines
indicate mean values.}
\end{figure}

\begin{figure}
\psfig{file=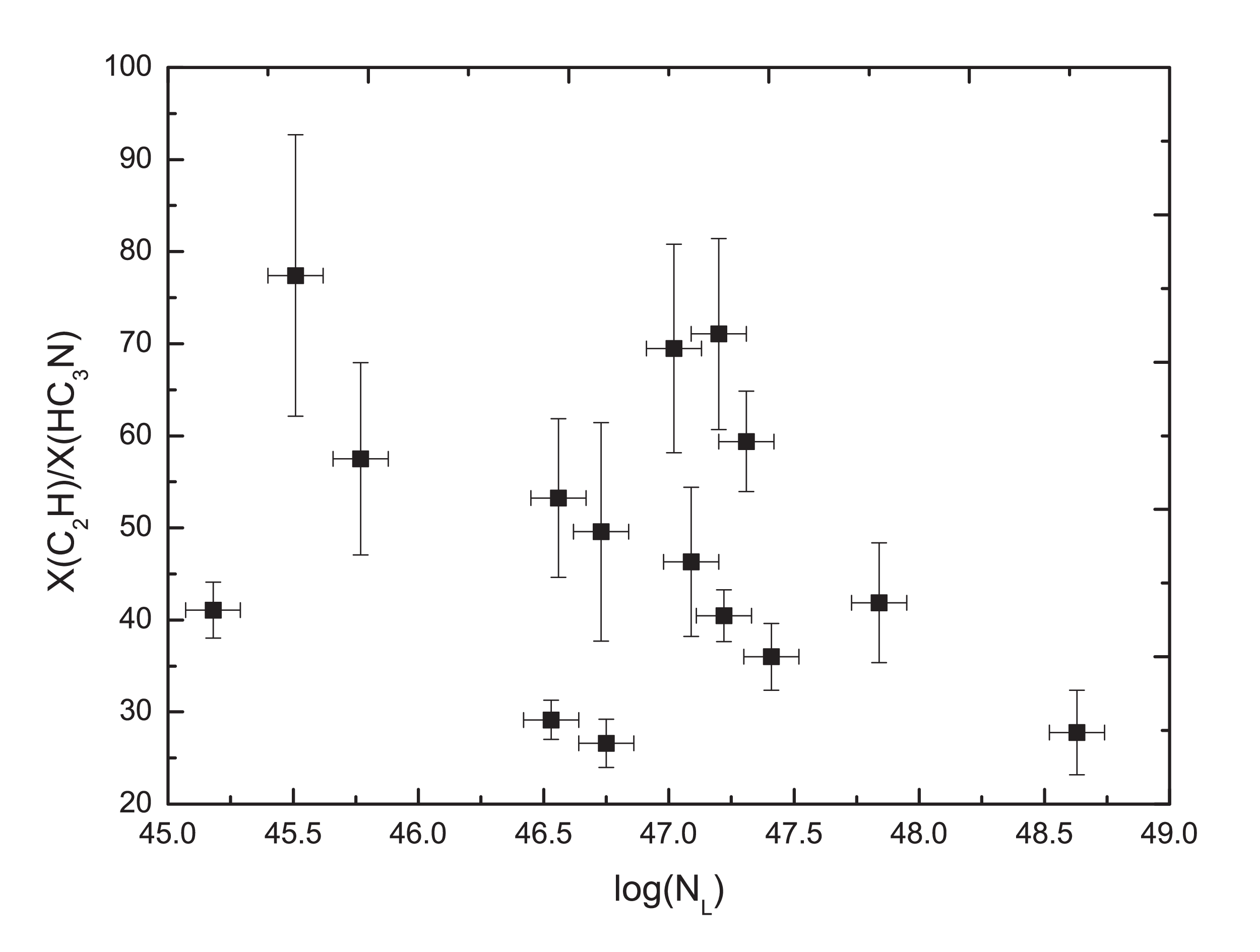,width=3.0in,height=2.4in}
 \caption{The [C$_2$H]/[HC$_3$N] relative abundance ratio plotted as a function of $N_L$ in logarithmic
scales.}
\end{figure}

\begin{figure}
\psfig{file=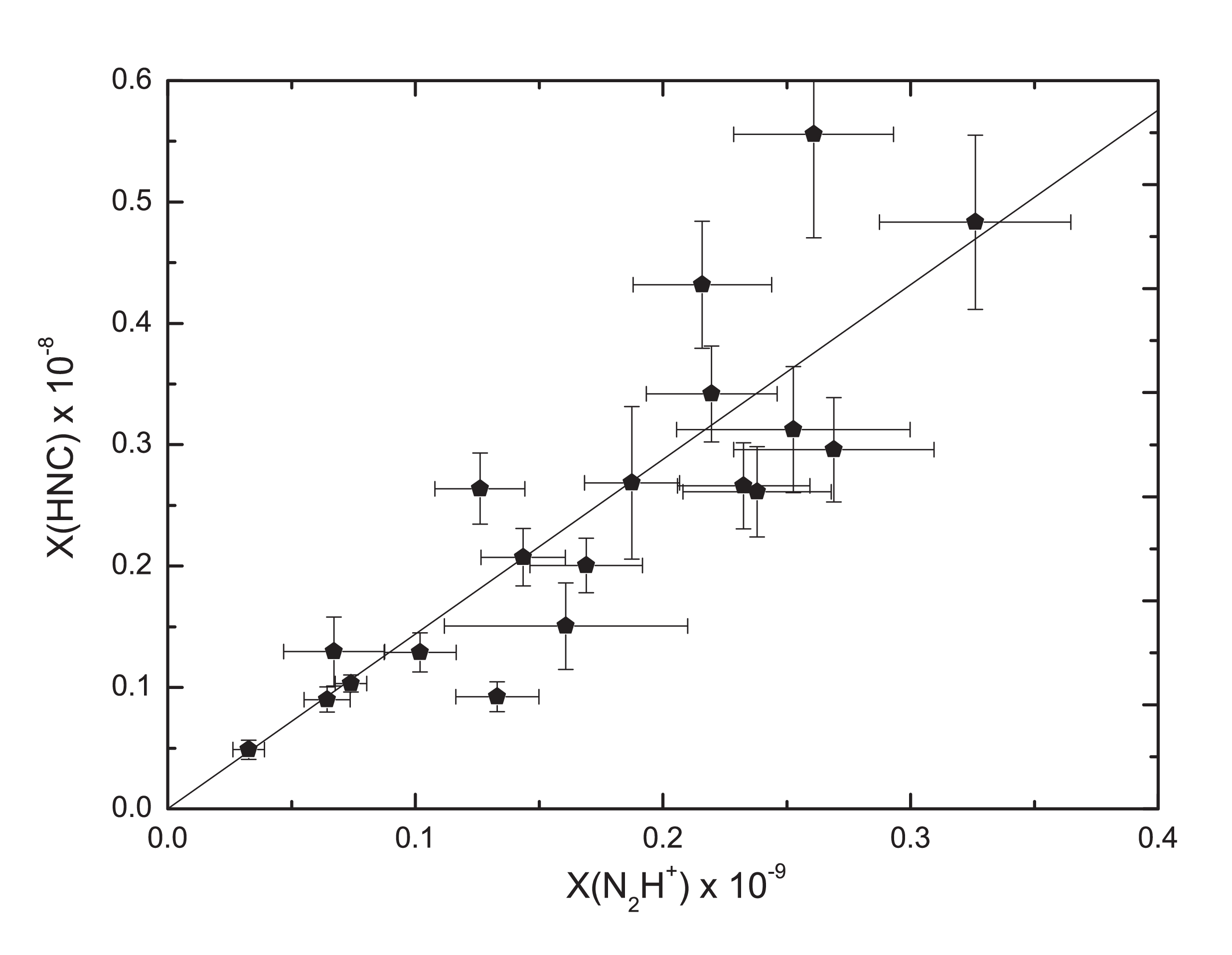,width=3.0in,height=2.4in}
 \caption{HNC fractional abundance plotted as a function of N$_2$H$^+$ abundance. The solid line shows the least square fit to the data.}
\end{figure}


\begin{thebibliography}{}
\bibitem[\protect\citeauthoryear{Contreras et al.}{2007a}]{b21} Belloche, A., Menten, K. M., Comito, C., et al. 2008, A\&A, 482, 179

\bibitem[\protect\citeauthoryear{Contreras et al.}{2007a}]{b21} Bergin, E. A., \& Langer, W. D. 1997, ApJ, 486, 316

\bibitem[\protect\citeauthoryear{Contreras et al.}{2007a}]{b21} Bergin, E. A., \& Tafalla, M. 2007, ARA\&A, 45, 339

\bibitem[\protect\citeauthoryear{Contreras et al.}{2007a}]{b21} Beuther, H., Semenov, D., Henning, T., \& Linz, H. 2008, ApJ, 675, L33

\bibitem[\protect\citeauthoryear{Contreras et al.}{2007a}]{b21} Casoli, F., Combes, F., Dupraz, C., Gerin, M., \& Boulanger, F. 1986, A\&A, 169,
281

\bibitem[\protect\citeauthoryear{Contreras et al.}{2007a}]{b21} Cesaroni, R., Hofner, P., Araya, E., \& Kurtz, S. 2010, A\&A, 509, 50

\bibitem[\protect\citeauthoryear{Contreras et al.}{2007a}]{b21} Cesaroni, R., Walmsley, C. M., \& Churchwell, E. 1992, A\&A, 256, 618

\bibitem[\protect\citeauthoryear{Contreras et al.}{2007a}]{b21} Chaisson, E. J., 1976, in Avrett E. H., ed., Frontiers of Astrophysics. Harvard
Univ. Press, Harvard, p. 259

\bibitem[\protect\citeauthoryear{Contreras et al.}{2007a}]{b21} Chapman, J. F., Millar, T. J., Wardle, M., Burton, M. G., \& Walsh, A. J.
2009, MNRAS, 394, 221

\bibitem[\protect\citeauthoryear{Contreras et al.}{2007a}]{b21} Chen, X., Gan, C.G., Ellingsen, S., et al. 2013, ApJS, 206, 22

\bibitem[\protect\citeauthoryear{Contreras et al.}{2007a}]{b21} Contreras, Y., Schuller, F., Urquhart, J. S., et al. 2013, A\&A, 549, A45

\bibitem[\protect\citeauthoryear{Contreras et al.}{2007a}]{b21} Daniel, F., Cernicharo, J., \& Dubernet, M.-L. 2006, ApJ, 648, 461

\bibitem[\protect\citeauthoryear{Contreras et al.}{2007a}]{b21} Dislaire, V., Hily-Blant, P., Faure, A., et al. 2012, A\&A, 537, 20A

\bibitem[\protect\citeauthoryear{Contreras et al.}{2007a}]{b21} Egan, M. P., Shipman, R. F., Price, S. D., et al.
1998, ApJ, 494, 199

\bibitem[\protect\citeauthoryear{Foster et al.}{2011}]{b4} Foster, J. B., Jackson, J. M., Barnes, P. J., et al. 2011, ApJS, 197, 25

\bibitem[\protect\citeauthoryear{Foster et al.}{2011}]{b4} Foster, J. B., Rathborne, J. M., Sanhueza, P., et al. 2013, PASA, 30, 38

\bibitem[\protect\citeauthoryear{Contreras et al.}{2007a}]{b21} Fuente, A., Martin-Pintado, J., Cernicharo, J., \& Bachiller, R. 1993, A\&A,
276, 473

\bibitem[\protect\citeauthoryear{Contreras et al.}{2007a}]{b21} Guzm\'{a}n, A., Garay, G.,  Brooks, K. J., \& Voronkov, M. A. 2012,
ApJ, 753, 51

\bibitem[\protect\citeauthoryear{Contreras et al.}{2007a}]{b21} Guzm\'{a}n, A., Sanhueza, P., Contreras, Y., et al. 2015, ApJ,
815, 130

\bibitem[\protect\citeauthoryear{Contreras et al.}{2007a}]{b21} Hennemann, M., Birkmann, S. M., Krause, O., et al. 2009, ApJ, 693,
1379

\bibitem[\protect\citeauthoryear{Contreras et al.}{2007a}]{b21} Henning, T., Pfau, W., \& Altenhoff, W. J. 1990, A\&A, 227, 542

\bibitem[\protect\citeauthoryear{Contreras et al.}{2007a}]{b21} Herbst, E., \& Klemperer, W. 1972, ApJ, 185, 505

\bibitem[\protect\citeauthoryear{Contreras et al.}{2007a}]{b21} Hirota, T., Yamamoto, S., Mikami, H., \& Ohishi, M. 1998, ApJ, 503,
717

\bibitem[\protect\citeauthoryear{Hoq et al. }{2013}]{b4.5} Hoq, S., Jackson, J.M., Foster, J.B., et al. 2013, ApJ, 777, 157

\bibitem[\protect\citeauthoryear{Hoq et al. }{2013}]{b4.5} Jackson, J. M., Rathborne, J. M., Foster, J. B., et al. 2013, PASA, 30, 57

\bibitem[\protect\citeauthoryear{Hoq et al. }{2013}]{b4.5} Kauffmann, J., Bertoldi, F., Bourke, T. L., Evans, N. J., II, \& Lee, C. W. 2008,
A\&A, 487, 993

\bibitem[Yu et al. (2013)]{abr08} Keto, E., \& Rybicki, G. 2010, ApJ, 716, 1315

\bibitem[\protect\citeauthoryear{Contreras et al.}{2007a}]{b21} Ladd, N., Purcell, C., Wong, T., \& Robertson, S. 2005, Publ. Astron. Soc.
Aust., 22, 62

\bibitem[\protect\citeauthoryear{Contreras et al.}{2007a}]{b21} Lee, J. E., Bergin, E. A., \& Evans, N. J. 2004, ApJ, 617, 360

\bibitem[\protect\citeauthoryear{Contreras et al.}{2007a}]{b21} Li, J., Wang, J. Z., Gu, Q. S., Zhang, Z. Y., \& Zheng, X. W. 2012, ApJ, 745, 47

\bibitem[\protect\citeauthoryear{Contreras et al.}{2007a}]{b21} Lumsden, S. L., Hoare, M. G., Urquhart, J. S. et al. 2013, ApJS, 208, 11.

\bibitem[\protect\citeauthoryear{Contreras et al.}{2007a}]{b21} Lumsden, S. L., Hoare, M. G., Oudmaijer, R. D., \& Richards, D.
2002, MNRAS, 336, 621.

\bibitem[\protect\citeauthoryear{Contreras et al.}{2007a}]{b21} Mauch, T., Murphy, T., Buttery, H. J., et al. 2003, MNRAS, 342, 1117

\bibitem[\protect\citeauthoryear{Lumsden et al.}{2007a}]{b21} M\"{u}ller, H. S. P., Thorwirth, S., Roth, D. A., \& Winnewisser, G.
2001, A\&A, 370, L49
\bibitem[\protect\citeauthoryear{Lumsden et al.}{2007a}]{b21} M\"{u}ller, H. S. P., Schl\"{o}der, F., Stutzki, J., \& Winnewisser, G. 2005, J. Molec. Struct., 742, 215

\bibitem[\protect\citeauthoryear{Lumsden et al.}{2007a}]{b21} Ossenkopf, V., \& Henning, T. 1994, A\&A, 291, 943

\bibitem[Yu et al. (2013)]{abr08} Pagani, L., Daniel, F., \& Dubernet, M.-L. 2009, A\&A, 494, 719

\bibitem[\protect\citeauthoryear{Contreras et al.}{2007a}]{b21} Perault, M., Omont, A., Simon, G., et al.
1996, A\&A, 315, 165

\bibitem[\protect\citeauthoryear{Contreras et al.}{2007a}]{b21} Peretto, N., \& Fuller, G. A. 2009, VizieR Online Data Catalog,
350, 50405

\bibitem[\protect\citeauthoryear{Contreras et al.}{2007a}]{b21} Pratap, P., Dickens, J. E., Snell, R L., et al. 1997, ApJ, 486, 862

\bibitem[\protect\citeauthoryear{Molinari et al.}{2007a}]{b21} Purcell, C. R., Balasubramanyam, R., Burton, M.G. et al. 2006, MNRAS, 367, 553

\bibitem[Yu et al. (2013)]{abr08} Purcell, C. R., Longmore, S. N., Burton, M. G., et al. 2009, MNRAS, 394, 323

\bibitem[\protect\citeauthoryear{Contreras et al.}{2007a}]{b21} Qin, S. L., Schilke, P., Rolffs, R., et al. 2011, A\&A, 530, L9

\bibitem[\protect\citeauthoryear{Contreras et al.}{2007a}]{b21} Rathborne, J. M., Longmore, S. N., Jackson, J. M., et al. 2014, ApJ, 786, 140

\bibitem[\protect\citeauthoryear{Molinari et al.}{2007a}]{b21} Sakai, T., Sakai, N., Hirota, T., \& Yamaguchi, N. 2010, ApJ, 714, 1658

\bibitem[\protect\citeauthoryear{Molinari et al.}{2007a}]{b21} Sanhueza, P., Jackson, J. M., Foster, J. B., et al. 2012, ApJ, 756, 60

\bibitem[\protect\citeauthoryear{Molinari et al.}{2007a}]{b21} Sanhueza, P., Jackson, J. M., Foster, J. B., et al. 2013, ApJ, 773, 123

\bibitem[\protect\citeauthoryear{Molinari et al.}{2007a}]{b21} Schuller F., Menten K. M., Contreras Y., Wyrowski F., Schilke P.,
Bronfman L., Henning T., et al. 2009, A\&A, 504, 415

\bibitem[\protect\citeauthoryear{Contreras et al.}{2007a}]{b21} Shematovich, V. I. 2012, SoSyR, 46, 391s

\bibitem[\protect\citeauthoryear{Contreras et al.}{2007a}]{b21} Sreenilayam, G., \& Fich, M. 2011, AJ, 142, 4

\bibitem[\protect\citeauthoryear{Siringo et al.}{2007a}]{b21} Tucker, K. D., Kutner, M. L., \& Thaddeus, P. 1974, ApJ, 193, L115

\bibitem[\protect\citeauthoryear{Contreras et al.}{2007a}]{b21} Urquhart, J. S., Busfield, A. L., Hoare, M. G., et al. 2007, A\&A, 461, 11

\bibitem[\protect\citeauthoryear{Contreras et al.}{2007a}]{b21} Urquhart, J. S., Busfield, A. L., Hoare, M. G., et al. 2008  A\&A, 487, 253

\bibitem[\protect\citeauthoryear{Urquhart et al.}{2007a}]{b21} Vasyunina, T., Linz, H., Henning, Th., et al. 2011, A\&A, 527, A88

\bibitem[\protect\citeauthoryear{Urquhart et al.}{2007a}]{b21} Vigren, E., Zhaunerchyk, V., Hamberg, M.,  et al. 2012, ApJ, 757, 34

\bibitem[\protect\citeauthoryear{Urquhart et al.}{2007a}]{b21} Yu, N. P., \& Wang, J. J. 2014, MNRAS, 440,
1213

\bibitem[\protect\citeauthoryear{Urquhart et al.}{2007a}]{b21} Yu, N. P., Wang, J. J., \& Li, N. 2014, MNRAS, 445,
3374

\bibitem[\protect\citeauthoryear{Urquhart et al.}{2007a}]{b21} Yu, N. P., \& Wang, J. J. 2015, MNRAS, 451, 2507

\bibitem[\protect\citeauthoryear{Urquhart et al.}{2007a}]{b21} Yu, N. P., Wang, J. J., \& Li, N. 2015, MNRAS, 446,
2566

\bibitem[\protect\citeauthoryear{Urquhart et al.}{2007a}]{b21} Watt, G. D., White, G. J., Millar, T. J., \& van Ardenne, A. 1988,
A\&A, 195, 257

\bibitem[\protect\citeauthoryear{Contreras et al.}{2007a}]{b21} Williams, J. P., \& Cieza, L. A. 2011, ARA\&A, 49, 67

\bibitem[\protect\citeauthoryear{Contreras et al.}{2007a}]{b21} Wirstrom, E. S., Geppert, W. D., Hjalmarson, A., et
al. 2011, Observational Tests of Interstellar Methanol Formation,
Astron. Astrophys., vol. 533, p. A24

\bibitem[\protect\citeauthoryear{Contreras et al.}{2007a}]{b21} Wu, L., Evans, N. J. II, Gao, Y., et al. 2005, ApJL, 635, 173

\bibitem[\protect\citeauthoryear{Contreras et al.}{2007a}]{b21} Wu, L., Evans, N. J. II, Shirley, Y.L., \& Knez, C. 2010, ApJS, 188, 313

\bibitem[\protect\citeauthoryear{Contreras et al.}{2007a}]{b21} Zinnecher, H., \& Yorke, H. W. 2007, ARA\&A, 45, 481

\end{thebibliography}
\end{document}